\documentclass[journal]{IEEEtran}
\IEEEoverridecommandlockouts

\usepackage{bbm}
\usepackage{graphicx}

\usepackage{xcolor}
\usepackage{float}   
\usepackage[normalem]{ulem}
\usepackage{dsfont}
\usepackage{textcomp}
\usepackage{stfloats}
\usepackage{url}
\usepackage{verbatim}
\usepackage{algorithmic}
\usepackage{array}
\usepackage{booktabs}
\usepackage{pifont}
\usepackage{comment}
\usepackage{algorithm}
\usepackage{algorithmic}
\usepackage[acronym]{glossaries}
\newacronym{ack}{ACK}{Acknowledgment}
\newacronym{agv}{AGV}{Automatic Guided Vehicles}
\newacronym{aoi}{AoI}{Age of Information}
\newacronym{aol}{AoL}{Age of Loop}
\newacronym{aoii}{AoII}{Age of Incorrect Information}
\newacronym{aos}{AoS}{Age of Synchronization}
\newacronym{ca}{CA}{Collision Avoidance}
\newacronym{coud}{CoUD}{Cost of Update Delay}
\newacronym{cowu}{CoWu}{Content-based Wake-up}
\newacronym{csma}{CSMA}{Carrier Sense Multiple Access}
\newacronym{drx}{DRX}{Discontinuous Reception}
\newacronym{id}{ID}{Identification}
\newacronym{idwu}{IDWu}{Identity-based Wake-up}
\newacronym{iiot}{IIoT}{Industrial Internet of Things}
\newacronym{if}{I/F}{interface}
\newacronym{iot}{IoT}{Internet of Things}
\newacronym{k-qaoi}{$k$-QAoI}{top-$k$ QAoI}
\newacronym{mac}{MAC}{Medium Access Control}
\newacronym{mec}{MEC}{Mobile Edge Computing}
\newacronym{ook}{OOK}{On-Off Keying}
\newacronym{osd}{OSD}{One-shot Data}
\newacronym{paoi}{PAoI}{Peak AoI}
\newacronym{qaoi}{QAoI}{Query Age of Information}
\newacronym{rr}{RR-scheduling}{round-robin scheduling}

\newacronym{tdma}{TDMA}{Time Division Multiple Access}
\newacronym{uav}{UAV}{Unmanned Aerial Vehicles}
\newacronym{voi}{VoI}{Value of Information}
\newacronym{wifi}{WiFi}{Wireless Fidelity}
\newacronym{wsn}{WSNs}{Wireless Sensor Networks}

\usepackage{amsmath,amsfonts, amssymb}
\usepackage{bm}
\usepackage{graphicx}

\usepackage[font=small]{caption}
\usepackage[font=footnotesize]{subcaption}
\usepackage{tikz}
\usetikzlibrary{plotmarks,patterns,decorations.pathreplacing,backgrounds,calc,arrows,arrows.meta,spy,matrix}
\usepackage{tikzscale}

\usepackage{pgfplots}
\usepgfplotslibrary{colorbrewer, groupplots, patchplots}
\pgfplotsset{
    compat=newest,
    legend style={font=\footnotesize, fill opacity=0.7,  draw opacity=1, text opacity=1, draw=white!15!black, legend cell align=left, align=left}, 
    width=0.8\columnwidth, 
    scale only axis,
    height=4cm,
    yminorticks=false,
    xminorticks=false,
    label style={font=\small},
    title style={font=\small},
    tick align=outside,
    tick pos=left,
    tick style={color=black},
    tick label style={font=\footnotesize},
    grid style={line width=.1pt, draw=gray!20},
    major grid style={line width=.1pt,draw=gray!20},
    plot coordinates/math parser=false 
}
\newlength\figureheight
\newlength\figurewidth

\usepackage{multirow}
\usepackage{tablefootnote}
\usepackage{booktabs}
\usepackage{tabularx}

\def\BibTeX{{\rm B\kern-.05em{\sc i\kern-.025em b}\kern-.08em T\kern-.1667em\lower.7ex\hbox{E}\kern-.125emX}}

\newcommand{\argmin}{\mathop{\rm arg~min}\limits}
\newcommand{\probP}{\text{I\kern-0.15em P}}

\definecolor{amaranth}{rgb}{0.9, 0.17, 0.31}

\begin{document}
%

\title{Content-based Wake-up for Energy-efficient and Timely Top-$k$ IoT Sensing Data Retrieval}
\author{Junya~Shiraishi,~\IEEEmembership{Member,~IEEE,}
        Anders~E.~Kal{\o}r,~\IEEEmembership{Member,~IEEE,}
        Israel~Leyva-Mayorga,~\IEEEmembership{Member,~IEEE,}
    Federico~Chiariotti,~\IEEEmembership{Senior Member,~IEEE,}
        Petar~Popovski,~\IEEEmembership{Fellow,~IEEE,}
   and~Hiroyuki~Yomo,~\IEEEmembership{Member,~IEEE}%
    \thanks{This work was partly supported by the Villum Investigator Grant ``WATER" from the Velux Foundation, Denmark, partly by the Horizon Europe SNS ``6G-XCEL" project with Grant 101139194, and partly by the Horizon Europe SNS ``6G-GOALS'' project with grant 101139232. The work of J. Shiraishi was supported by Horizon Europe Marie Sk{\l}odowska-Curie Action (MSCA) Postdoc Fellowships with grant No~101151067. The work of A. E. Kal{\o}r was supported by the Independent Research Fund Denmark (IRFD) under Grant 1056-00006B. The work of H. Yomo was supported by JSPS KAKENHI under Grant JP22K04114. The work of F. Chiariotti was supported by the NextGenerationEU project ``REDIAL,'' under the Italian NRRP ``Young Researchers'' scheme (SoE0000009). (Corresponding author is Junya Shiraishi)}
   \thanks{J. Shiraishi, I. Leyva-Mayorga, and P. Popovski are with the Department of Electronic Systems, Aalborg University, 9220 Aalborg, Denmark (e-mail: jush@es.aau.dk; ilm@es.aau.dk; petarp@es.aau.dk). A. E. Kal{\o}r is  with the Faculty of Science and Technology, Keio University, Yokohama 223-8522, Japan and with the Department of Electronic Systems, Aalborg University, 9220 Aalborg, Denmark (e-mail: aek@keio.jp). F. Chiariotti is with the Department of Information Engineering, University of Padova, 35131 Padua, Italy (e-mail: chiariot@dei.unipd.it). H. Yomo is with the Graduate School of Science and Engineering, Kansai University, 564-8680 Suita, Japan (e-mail: yomo@kansai-u.ac.jp).}
   }

\maketitle

\begin{abstract} 
Energy efficiency and information freshness are key requirements for sensor nodes serving Industrial Internet of Things (IIoT) applications, where a sink node collects informative and fresh data before a deadline, e.g., to control an external actuator. Content-based wake-up (CoWu) activates a subset of nodes that hold data relevant for the sink's goal, thereby offering an energy-efficient way to attain objectives related to information freshness. This paper focuses on a scenario where the sink collects fresh information on top-$k$ values, defined as data from the nodes observing the $k$ highest readings at the deadline.  
We introduce a new metric called top-$k$ Query Age of Information ($k$-QAoI), which allows us to characterize the performance of CoWu by considering the characteristics of the physical process. Further, we show how to select the CoWu parameters, such as its timing and threshold, to attain both information freshness and energy efficiency. The numerical results reveal the effectiveness of the CoWu approach, which is able to collect top-$k$ data with higher energy efficiency while reducing $k$-QAoI when compared to round-robin scheduling, especially when the number of nodes is large and the required size of $k$ is small. 
\end{abstract}


%
\IEEEpeerreviewmaketitle

\section{Introduction}\label{sec:Intro}
\IEEEPARstart{L}{ow}-power \gls{wsn} support massive \gls{iot} deployments, including the emerging \gls{iiot} applications~\cite{sisinni2018industrial,mao2021energy}, in which the collected data are used for remote control/monitoring. As most sensor nodes in \gls{wsn} are powered by batteries, high energy efficiency of sensor nodes is a key requirement in \gls{wsn}. A promising technology to achieve this is wake-up radio, which enables each sensor node receive wake-up requests from the sink node using an ultra-low power wake-up receiver, while keeping its primary radio \gls{if} switched off~\cite{IEICE-yomo,piyare2017ultra}. This can greatly reduce the energy consumption of sensor nodes during idle periods, while allowing pull-based (demand-driven) data collection through wake-up signaling~\cite{IEICE-yomo,TGCN_Content}. For this reason, wake-up signaling is currently being considered in standardization bodies, such as 3GPP~\cite{3GPP_standard,wagner2023low}. Traditionally, wake-up signaling has been used to wake up specific nodes based on unique node or group identifiers~\cite{IEICE-yomo}. However, the wake-up signal can also be used to wake up nodes dynamically, e.g., based on their current sensor readings. This is the core idea in \gls{cowu}~\cite{TGCN_Content,shiraishi2022query}, where the wake-up signal is used to encode a threshold, such that only the nodes that observe values greater than the threshold are activated. Specifically, if the observed value of sensor node is equal to or larger than the transmitted threshold, the node activates its main radio \gls{if} and transmits data to the sink, and otherwise stays in a sleep state. By carefully controlling the wake-up threshold, it is possible to approximate a large number of data queries, such as retrieving the $k$ largest sensor readings, known as the top-$k$ query~\cite{TGCN_Content}, and retrieving sensor readings that fall within a certain range, i.e., a range query~\cite{shiraishi2022query}.

Among the possible queries, the top-$k$ query is one of the most common queries in \gls{wsn}, due to its many applications, e.~g., environmental monitoring/control~\cite{EXTOK}. More generally, the sensors may compute a certain score related to the query, such as similarity, and the intention is to collect the data from the $k$ sensors that have the highest scores~\cite{kalor2023random}. In most cases the data is collected to perform an action, such as controlling actuators or producing an alarm. 
Furthermore, such an action may often be associated with a strict time constraint, forcing the sink to obtain the top-$k$ set before a given deadline. For example, collecting the $k$ largest values with high information freshness at the deadline is important to monitor processes and identify outliers in factory scenarios or for digital twins~\cite{shiraishi2022query}. In the context of natural disaster management, such as wildfire and chemical gas leakage detection, a timely top-$k$ query is important to identify the most critical locations to, e.g., trigger interventions. However, the unreliable nature of the communication channel and the fact that the physical process monitored by the sensors change over time introduce a trade-off between the reliability and freshness of the collected data at the deadline through the timing of sampling, data transmission, and query transmission. Specifically, if the sink requests the top-$k$ set long before the deadline, the probability that it receives all $k$ observations is large, even in the event of packet losses that necessitate retransmissions. However, the physical process may change between the query instant and the deadline, resulting in a potentially large discrepancy between the measurements received by the sink and the values of the physical processes at the deadline. On the other hand, if the sink transmits the top-$k$ query close to the deadline, then there may not be sufficient time for retransmissions but the retrieved values are more likely to remain the same until the deadline.
The goal of this paper is to realize both timely and energy-efficient top-$k$ data collection by exploiting wake-up radio technology.

\begin{figure}[t]
\centering
\includegraphics[width=0.48\textwidth]{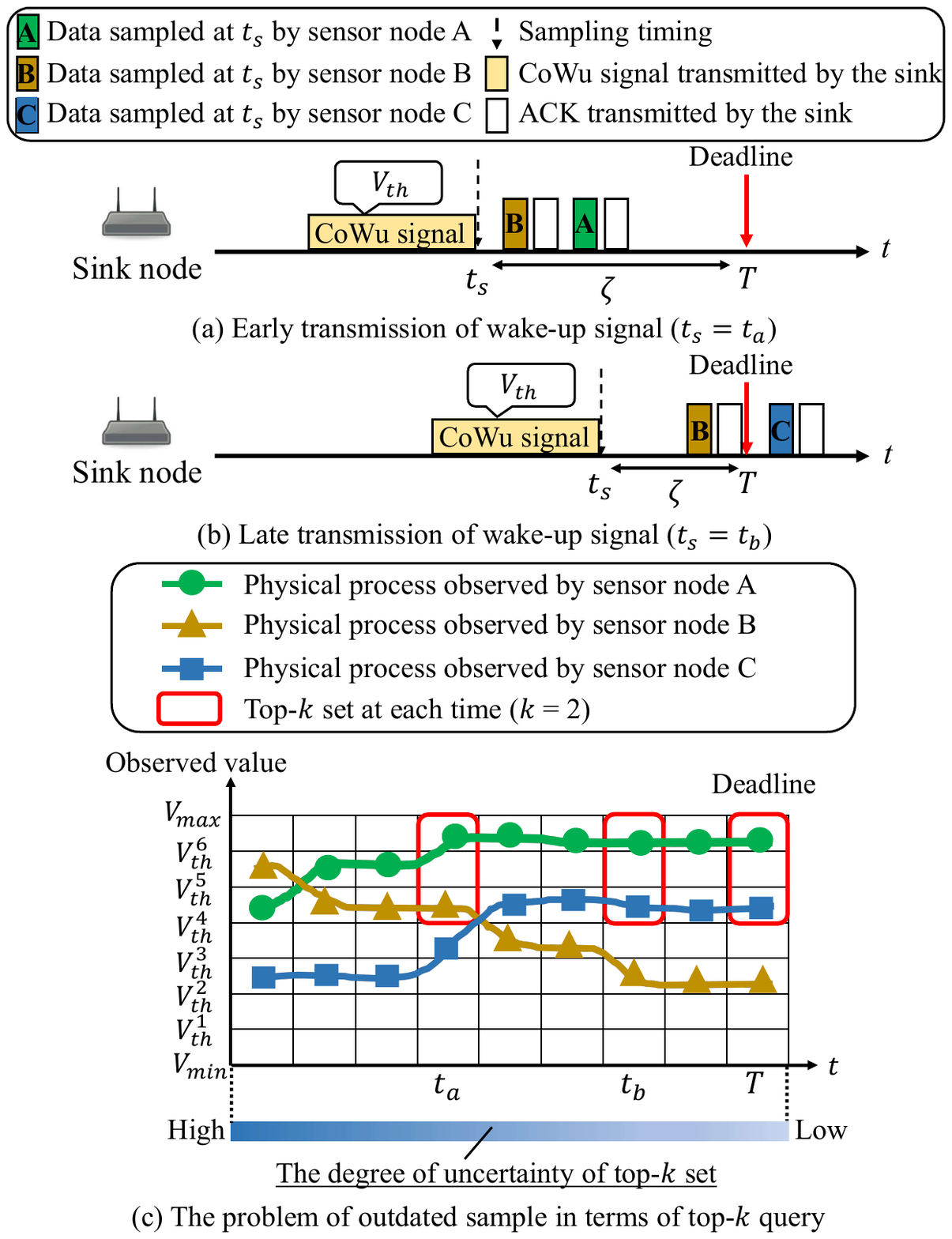}
\caption{An example of the top-$k$ query employing \gls{cowu}, where $k = 2$. (a): The sink transmits \gls{cowu} signal at the early time against the deadline. (b): The sink transmits \gls{cowu} signal at the late time against the deadline. (c) An example on how the timing and the threshold of \gls{cowu} signal affects the accuracy of a top-$k$ set at the deadline.}
\label{Fig:Data_Collection_Model_CoWu}
\end{figure}

\subsection{Content-based Wake-up for Timely Top-$k$ Query}
As argued earlier and also indicated in~\cite{shiraishi2022query}, sending a \gls{cowu} signal too early increases the risk that the collected values will be outdated at the deadline, while sending it too late may prevent some of the sensors from successfully transmitting their packets before the deadline. This is illustrated in Fig.~\ref{Fig:Data_Collection_Model_CoWu}(a)-(b), where a threshold $V_{\mathrm{th}}$ is transmitted using wake-up signalling $\zeta$ time slots prior to the deadline in order to collect the $k$ highest readings. In Fig.~\ref{Fig:Data_Collection_Model_CoWu}(a), the query is transmitted at an early time (large $\zeta$), resulting in the retrieval of all observations but with a high probability of being outdated at the deadline. On the other hand, in Fig.~\ref{Fig:Data_Collection_Model_CoWu}(b) the query is transmitted at a late time (small $\zeta$), resulting in only retrieving one observation, but with a low probability of being outdated at the deadline.

The impact of $\zeta$ and $V_{\mathrm{th}}$ on the accuracy at the deadline $T$ is also exemplified in Fig.~\ref{Fig:Data_Collection_Model_CoWu}(c) for $k = 2$. Here, $t_s=t_{a}$ and $t_s=t_{b}$ in Fig.~\ref{Fig:Data_Collection_Model_CoWu}(c) represent the case of the early transmission of wake-up signal (c.f. Fig.~\ref{Fig:Data_Collection_Model_CoWu}(a)) and late transmission of wake-up signal (c.f. Fig.~\ref{Fig:Data_Collection_Model_CoWu}(b)), respectively. The example shown in Fig.~\ref{Fig:Data_Collection_Model_CoWu}(c) highlights that the top-2 set at $T$, i.e., the true top-$k$ set, is $\{A, C\}$, while the nodes belonging to the estimated top-$k$ set deviate from the true top-$k$ set as the time difference between the latest update and $T$ becomes larger. Specifically, in the example of Fig.~\ref{Fig:Data_Collection_Model_CoWu}(c), the top-$k$ set at time slot $t_{b}$ is the same as the one at the deadline $T$ because $\zeta = T - t_{b}$ is small. On the other hand, the top-$k$ set at $t_{a}$ differs from the one at time $T$ because $\zeta = T - t_{a}$ is large. In general, the probability that the data that belongs to the top-$k$ set at the sampling time are equal to the data that belongs to the top-$k$ set at the deadline $T$ increases as the difference between sampling time and deadline decreases, and vice-versa. From Fig.~\ref{Fig:Data_Collection_Model_CoWu}(c), we can also observe the importance of the threshold of \gls{cowu}. If we set a high threshold, such as $V_{\mathrm{th}}^{6}$, the expected number of wake-up nodes is small enough to avoid collisions and ensure reception, but it might be too small to activate all the nodes belonging to the top-$k$ set. Conversely, setting a low threshold, such as $V_{\mathrm{th}}^{3}$, increases the probability that all the nodes belonging to the top-$k$ set are successfully activated, but can potentially lead to congestion in the access network, resulting in the failure of data collection by the deadline. 

To characterize the trade-off between successful transmissions and the freshness of the top-$k$ set at the deadline, in this paper we introduce a new metric called \gls{k-qaoi}, which we describe in detail in Sec.~\ref{sec:intro_k_QAoI}. Building upon the \gls{aoi}~\cite{kaul2011minimizing,kaul2012real} and \gls{qaoi}~\cite{chiariotti2022query} metrics, \gls{k-qaoi} characterizes the freshness of the top-$k$ set that the sink node has retrieved at the deadline. The \gls{k-qaoi} enables tractable analysis of the timeliness of the top-$k$ query using wake-up radio, and thus allows us to evaluate and compare the different energy-efficient data collection methods. The \gls{k-qaoi} metric has three features, namely, non-linear age, age penalty for data transmission failure, and quantification of the freshness of top-$k$ set, as summarized below:
\begin{itemize}
\item{Feature 1 [Non-linear age]: The freshness of received data depends on the time elapsed since the packet is generated in accordance with the rate of changes of physical process. In order to take this into account, we introduce two types of \emph{cost functions} for different types of physical processes with different rates of change.}
\item{Feature 2 [Age penalty for data transmission failure]: The sink can retrieve the information on the physical process observed by sensor nodes at the data generation time only when their data are successfully received. If it fails, the information freshness the sink holds becomes lower, which depends on the time elapsed since the last received update. We reflect the staleness caused by the failure of data transmission by introducing the age penalty.} 
\item{Feature 3 [Quantification of the freshness of top-$k$ set the sink obtains at the deadline]: Although the metric \gls{k-qaoi} itself cannot track the ground truth on the current top-$k$ set due to the analytical difficulty of calculating all possible ranking variations from the sampling timing to the deadline timing, it enables us to evaluate the freshness of collected top-$k$ set at the deadline timing. The basic idea comes from the simple observation that the possibility that the true top-$k$ set, i.e., the top-$k$ set at the deadline $T$, deviates from the one at the sampling timing increases as the duration between sampling time and deadline time becomes larger. The metric \gls{k-qaoi} introduced in this paper can capture this aspect by adopting the idea mentioned in features 1 and 2.}
\end{itemize}

\subsection{Related Work}

\subsubsection{Age of Information and Timing Requirements}
\gls{aoi}~\cite{kaul2011minimizing,kosta2017age} is a metric that quantitatively measures the freshness of information instead of the conventional throughput and latency. 
Plenty of information freshness metrics suited for specific applications/scenarios have been investigated. \gls{paoi}~\cite{costa2014age} is an \gls{aoi} variant that was introduced to track the worst-case age in packet management systems. In communication and control systems, information freshness metrics relating to specific applications have been actively studied and introduced, such as  \gls{aoii}~\cite{maatouk2020age}, \gls{voi}~\cite{alawad2022value,ayan2019age}, \gls{coud}~\cite{kosta2020cost}, \gls{aos}~\cite{tang2020scheduling}, Age of Changed Information~\cite{wang2021age}, and Age of Correlated Information~\cite{yin2020application,he2019joint}.  \gls{aoii} considers the penalty of the estimation error at the destination along with the time penalty in monitoring application. \gls{voi} is defined as a measure of uncertainty reduction from the information set of the receiver if the transmission is successful~\cite{ayan2019age}. In~\cite{kosta2020cost}, the authors introduced so-called \gls{coud}, which considers a non-linear aging process, while \gls{aos} evaluates desynchronization of information between the receiver and source. Age of Changed Information considers not only the time lag of the update, but also changes in the content of these updates. Age of Correlated Information considers the application level requirement, in which the information of the source can be updated if the correlated data generated at the same time is successfully collected.

\gls{aoi} also has been introduced in \gls{wsn}-related research~\cite{moltafet2021power}. For example, in~\cite{liu2020uav,gao2023aoi}, the authors introduced \gls{aoi} in the scenario of data collection using \gls{uav}, and in~\cite{wang2021sleep}, the authors introduced \gls{aoi} into sleep-wake sensors in an \gls{iiot} scenario, in which the sleep time can be set based on the physical processes and the importance of sensors with age-penalty function. In~\cite{zhang2022aoi}, authors introduced \gls{aoi} in \gls{wsn} and \gls{mec}. 

Most works related to \gls{aoi} assume a push-based communication model, in which a node generating data autonomously decides whether and when to transmit an update to a data collection node. However, these push-based metrics do not fully apply to pull-based communication, in which the communication protocol should be designed considering the generating process of the requesting application. In a pull-based communication regime, the users’ interest is in the freshness of information at the specific time instant when they make a query or decision~\cite{li2020waiting,chiariotti2022query}. 
The \gls{aoi} of pull-based transmission strategies has previously been studied in~\cite{chiariotti2022query} and the related \gls{voi} metric was analyzed in~\cite{chiariotti2022scheduling}. In~\cite{holm2023goal}, we also proposed a framework of query-aware sensor scheduling, in which the sensor with the highest likelihood of having an informative update is selected by the edge node in accordance with the query function. Further, considering the closed-loop of wireless network control system, the concept of \gls{aol} has been introduced~\cite{de2021age}, which can capture the information freshness of both downlink and uplink communication.

\subsubsection{Energy-saving for \gls{wsn}}
The basic approach to improve the energy efficiency of sensor nodes operating with batteries is duty-cycling~\cite{DC-Survey}, by which the switches of main radio are periodically turned on/off. 
However, it is difficult to satisfy the requirement of both lower latency and high energy efficiency. Wake-up radio~\cite{IEICE-yomo,piyare2017ultra,WakeupR}, which can be realized by installing the low-power wake-up receiver into sensor nodes, represents a promising approach to save wasteful energy consumption of sensor nodes, especially during an idle period. According to~\cite{piyare2017ultra}, the study of the wake-up receiver can be divided into hardware design or \gls{mac} and routing protocol design. The focus of this paper is the \gls{mac} protocol design for \gls{wsn} employing wake-up receivers.  

The most common wake-up control is \gls{idwu}, which aims to activate nodes based on a predetermined \gls{id}~\cite{piyare2017ultra,3GPP_standard}. 
Another type of wake-up signaling is \gls{cowu}~\cite{TGCN_Content}, which realizes the activation of nodes based on the content observed by each node. \gls{cowu} is suitable for creating clusters based on sensors' readings~\cite{ClusteringWakeupreceiver} and for the applications where the sink is interested in specific types of data, such as anomaly detection~\cite{kawakita2022wake}, identifying multiple emission sources~\cite{shiraishi2023wake}, range-query~\cite{shiraishi2022query}, and top-$k$ query~\cite{TGCN_Content,Multihop_CoWu,shiraishi2021periodical}. 
In~\cite{shiraishi2022query}, we introduced \gls{cowu} for the timely range query scenario, where we clarified the importance of the timing of \gls{cowu} signaling with respect to the deadline to realize higher accuracy of the retrieved range set. Compared to the top-$k$ query considered in this paper, the range query is simpler to implement as it does not require knowledge of the data distribution. In~\cite{TGCN_Content}, we introduced \gls{cowu} for top-$k$ monitoring scenario and proposed a data collection algorithm applying \gls{cowu}, which we extended to the multi-hop scenario in~\cite{Multihop_CoWu}. 
In~\cite{shiraishi2021periodical}, we exploited the temporal correlation of observed data to realize both high energy efficiency and high-ranking accuracy. In~\cite{yomo2021wake}, we applied Kernel density estimation methods to the top-$k$ query scenario to satisfy the trade-off between data collection delay and total energy consumption. However, all of these works ignored the aspect of freshness, which is crucial in, e.g., control scenarios in \gls{wsn}. How to collect top-$k$ set with high energy efficiency and high information freshness by the deadline has not been investigated in these related studies. In this paper, we go beyond our previous work~\cite{TGCN_Content,shiraishi2022query} by designing the wake-up control that realizes the timely and energy-efficient top-$k$ query retrieval, in which we investigate the optimal threshold and timing of \gls{cowu} signal against the different types of physical process, etc., leveraging the introduced metric, \gls{k-qaoi}.

\subsection{Contributions and Paper Structure}
The contributions can be summarized as follows:
\begin{itemize}
\item{This article introduces a new metric called \gls{k-qaoi} for top-$k$ queries with respect to a given deadline, enabling a tractable analysis of the data freshness at the deadline. }
\item{We characterize the optimal timing of wake-up signaling against the rate of change of the physical process and the temporal knowledge of the sink as a function of age penalty, which gives us insight into the design of wake-up control in a timely data collection scenario.}
\item{We conduct a thorough theoretical analysis under the assumption of a realistic \gls{mac} protocol to determine the threshold of \gls{cowu} and the timing of \gls{cowu} signaling to retrieve the fresh top-$k$ set by the deadline. We further approximately obtain the optimized set of parameters for \gls{cowu} signaling that minimizes total energy consumption while satisfying the constraint of \gls{k-qaoi}.}
\item{The paper presents an extensive performance evaluation of \gls{cowu} and comparison against the three baselines, \gls{rr}, probability $q$-based Random Wake-up, and Genie-aided scheme. The evaluation considers practical assumptions, such as packet error and different characteristics of physical processes. The results show the region in which \gls{cowu} shows superiority against \gls{rr} in terms of total energy consumption and \gls{k-qaoi}.}
\end{itemize}

The remainder of the paper is organized as follows. In Sec.~\ref{sec:sys} we present the system model and introduce \gls{k-qaoi}. We analyze the performance of the considered scheme in Sec.~\ref{sec:Theoretical_Analysis} by deriving the equations on the \gls{k-qaoi} and energy consumption of nodes and formulate optimization problem and its solution in Sec.~\ref{sec:Optimization}. Then, we present our results in Sec.~\ref{sec:sim}. Finally, we conclude the article in Sec.~\ref{Sec:conc} and discuss possible avenues for future work.

\section{System Model}\label{sec:sys}
\subsection{Scenario}\label{sec:scenario}
We study a scenario comprising a sink node and $N$ sensor nodes equipped with wake-up receivers. Each sensor, indexed by $n=1,\ldots,N$, monitors a
random process $V_n(t)\in [V_{\mathrm{min}}, V_{\mathrm{max}}]$, whose time samples are independent and identically distributed. The time is divided into recurring episodes, and each episode is divided into slots. The episodes are assumed to occur at random and infrequent instants, so that the processes monitored by the sensors can be assumed to be independent between episodes. The beginning of each episode is triggered by a request to collect data from the sensor nodes, which must be gathered before a given deadline $T$, as illustrated in Fig.~\ref{Fig:Data_Collection_Model_CoWu}. The requests originate from an external entity, such as an actuator or monitoring software. In this paper, we focus on top-$k$ queries, i.e., the sink aims to retrieve values of the $k$ largest readings ($k \ll N$). Given the dynamic nature of the monitored processes, collecting the top-$k$ set, i.e., the $k$ largest readings, with high freshness at the deadline is important. Note that the value of $k$ depends on the users' requirement, whose size depends on the specific applications, and may change between episodes.   
Here, we focus on the specific instance of a timely top-$k$ query, in which the sink knows the size of $k$ before starting the data collection.

\subsection{Transmission Model}
\subsubsection{Wake-up signal}
We consider a scheme where the sink transmits wake-up signal using its main radio interface (i.e., no special hardware installation is required)~\cite{IEICE-yomo, TGCN_Content}. Here, we assume that the wake-up receiver operates over the same frequency band as the main radio. This makes it possible for each wake-up receiver to detect signals transmitted by the main radio of the sink. 
The devices are equipped with a simple wake-up receiver, which conducts non-coherent envelope detection and \gls{ook}. With this operation, the wake-up receiver only extracts the information of length of each frame (without decoding the information within header/payload), to which the information on wake-up signaling is embedded~\cite{IEICE-yomo, TGCN_Content}. The detailed configuration of the wake-up receiver can be found in~\cite{IEICE-yomo, TGCN_Content}. During sleep, each node switches off its main radio interface and activates the wake-up receiver to monitor the communication requests. When the sink receives the communication requests, it transmits a wake-up signal to activate the sleeping target sensor nodes at a specific time considering the deadline. The wake-up receiver detecting the wake-up signal activates its main radio and transmits its data following the data transmission model described in Sec.~\ref{sec:data_transmission}, if it satisfies the wake-up condition notified through frame-length based signaling.

As a wake-up control, we focus on \gls{cowu}~\cite{TGCN_Content}, in which the threshold $V_{\mathrm{th}}$ is embedded into a wake-up signal which is transmitted to the wake-up receivers through a secondary, low-power radio link. This is achieved by encoding the estimated threshold of the top-$k$ set $V_{\mathrm{th}}$ into the wake-up signal, for example, using \gls{ook}~\cite{TGCN_Content}. 
For example, the information on the threshold can be encoded into different frame lengths, in which one can map the higher (lower) threshold $V_{\mathrm{th}}$ to the shorter (longer) frame length $T_{\mathrm{wu}}$ as in~\cite{TGCN_Content}. This frame length can be expressed as $T_{\mathrm{wu}}(V_{\mathrm{th}}) = T_{\mathrm{min}} + \mathcal{I}(V_{\mathrm{th}})T_{\mathrm{step}}$. Here, the $T_{\mathrm{min}}$ is the minimum frame, $T_{\mathrm{step}}$ is the value of the step of the different frame lengths, and $\mathcal{I}(V_{\mathrm{th}})$ is the quantization interval of threshold $V_{\mathrm{th}}$, where higher $V_{\mathrm{th}}$ corresponds to the smaller quantization interval. Note that the maximum number of bits that can be encoded into the wake-up signal is limited, such as $2^9$ = 512 due to the practical constraint. The possible frame length for the wake-up signal can be set by considering the standard IEEE 802.15.4g and the data size used for the data transmission~\cite{TGCN_Content}. Further, $T_{\mathrm{step}}$ can be chosen considering the trade-off between the resolution and the false wake-up ratio~\cite{shiraishi2021periodical}.

Following the reception of a collection request, the sink node picks a time slot at time $t_s$, which is $\zeta \in \mathbb{Z}^{+}$ slots earlier than the deadline time $T$, and broadcasts a top-$k$ query to the sensors using \gls{cowu} so that it can be detected by wake-up receivers at time slot $t_s$\footnote{For simplicity, this paper focuses on the impact of the timing of wake-up signaling reception $t_{s}$ at the wake-up receiver. In practice, the sink should tune the timing of wake-up signal transmission by considering the length of the \gls{cowu} signal, which depends on the content and propagation delay.}. We will assume that each node samples its data immediately after receiving the \gls{cowu} signal as in \cite{shiraishi2022query}, for the sake of simplicity. We defer alternative sampling policies/wake-up control considering characteristics of the physical process observed by each sensor to future work. We further assume that the values of the monitored processes are drawn independently from a common distribution $g(v;t_{s})$ at the sampling instant\footnote{This can be extended to the multi-attribute setting, where the sink requests multiple attributes of top-$k$ data by the deadline. The design and analysis of wake-up control for such a setting are kept for future work.}, i.e., $V_n(t_s)\overset{\text{iid}}{\sim}g(v;t_{s})$. 
Since the physical processes evolve in the $\zeta$ steps between the sampling time slot, $t_s$, and the deadline, $T$, our goal is to collect top-$k$ data that \emph{remain timely} at the deadline, as formalized later in Sec.~\ref{sec:intro_k_QAoI}.
In \gls{cowu}, the wake-up receiver detecting the wake-up signal compares its observations $V_{\mathrm{o}}$ with the threshold $V_{\mathrm{th}}$. If the observation is higher than the threshold, i.e., $V_{\mathrm{o}} \geq V_{\mathrm{th}}$, it activates its main radio and transmits data following the data transmission model described in Sec.~\ref{sec:data_transmission}.

\subsubsection{Data transmission}\label{sec:data_transmission}
The sensor nodes that wake up after receiving the wake-up signal attempt data transmission through the primary radio over unlicensed frequency bands, e.g., following the IEEE 802.15.4 standard~\cite{IEEE15.4}, and this paper assumes the use of a $p$-persistent \gls{csma} protocol for transmitting observations. Note that, as shown in~\cite{epoch}, a $p$-persistent \gls{csma} protocol is suited for approximating the practical \gls{mac} protocol of IEEE 802.15.4, as it exploits both the \gls{csma} and \gls{ca} mechanisms~\cite{TGCN_Content}. Therefore, each node with a relevant packet conducts carrier sensing at the beginning of a slot. Here, the duration of each slot is denoted as $\delta$ [s]. If the channel is idle, the node attempts transmission of its packet with transmission probability $p$. Each transmission attempt occupies $L$ consecutive time slots. We consider a collision channel with erasures, where collisions cause packet loss with probability $1$. In the absence of a collision, erasures occur with probability $e_{c}$, which is the same for all devices. An error-free \gls{ack} is transmitted from the sink after each successful sensor data transmission. The nodes receiving an \gls{ack} transition back to sleep-state, while the nodes detecting a packet loss by the absence of an \gls{ack} retransmit their packets following $p$-persistent \gls{csma}. Developing mechanisms to mitigate the hidden terminal problem is out of the scope of this paper and, hence, we assume that all nodes, including the sink, are located within each other's communication, wake-up, and carrier-sensing range.

\subsection{Energy Model}\label{sec:energy_model}
When there is no communication request, each sensor node consumes power only due to the wake-up receiver, which is negligible. The devices that are activated by the wake-up signal consume power either by transmitting or receiving data from the sink. We denote the power consumption in the transmission and receiving states by $\xi_{T}$ and $\xi_{R}$, respectively. The total energy consumption of all sensor nodes over an episode is
\begin{equation}
\ E_{\mathrm{total}} (N)=\sum_{n = 1}^N\xi_{T}t_{\mathrm{tx}}^n + \xi_{R}t_{\mathrm{rx}}^n,\label{eq:Energy model}
\end{equation}
where $t_{\mathrm{tx}}^n$ and $t_{\mathrm{rx}}^n$ are time spent in transmission and receiving state for the sensor node $n$, respectively, within the episode. As described in Sec.~\ref{sec:data_transmission}, each wake-up node transmits data based on $p$-persistent \gls{csma} and transitions back to the sleep state when it receives \gls{ack}. This operation leads to the different values of $t_{\mathrm{tx}}^n$ and $t_{\mathrm{rx}}^n$ for all nodes. Note that if the sensor node $n$ remains in the sleep state during the whole episode, $t_{\mathrm{tx}}^n =0$ and $t_{\mathrm{rx}}^n =0$.

\subsection{The \gls{k-qaoi} Metric}\label{sec:intro_k_QAoI}
To characterize the freshness of the retrieved top-$k$ set at the deadline, we introduce a new metric called \gls{k-qaoi}, which builds upon the \gls{coud} metric~\cite{kosta2020cost}. Recall that the classical \gls{aoi}~\cite{kosta2017age} is defined as the time elapsed since the last received packet was generated. Formally, let us denote the time slot at which $j$-th observation value of sensor node $n$ is successfully received at the sink node as $t'_{n,j}$ and the time slot this successfully received packet is generated at the application layer as $t_{n,j}$. Then, in time slot $t$, let $u_{n}(t)$ be the data generation time slot of the most recently received packet from sensor node $n$, which can be described as follows:
\begin{equation}
u_{n}(t)=\max_{j: t'_{n,j}{\leq}t}t_{n,j}. 
\end{equation}
The \gls{coud} of node $n$ at time slot $t$ is then defined as
\begin{equation}
\ \Delta_{n}(t)=f_s\left(t-u_{n}(t)\right),\label{eq:AoI_with_CoUD}
\end{equation}
where $f_s(\cdot)$ is a non-decreasing cost function that describes the cost of having an age of $t-u_{n}(t)$, e.g., reflecting the dynamics of the physical process monitored by the node. In this work, we restrict the focus to two cost functions, namely linear age and exponential age, defined as
\begin{equation}
\ f_{s}(\tau)=\begin{cases}
\tau,&\text{linear age},\\
e^{{\alpha}\tau}-1,&\text{exponential age},
\end{cases}\label{eq:cost_function}
\end{equation}
for some positive constant $\alpha$.
Note that the linear age corresponds to the original \gls{aoi} metric, while the exponential age corresponds to the scenario where the autocorrelation of the physical process is small~\cite{kosta2020cost}.

In this paper, we are concerned about the \gls{coud} only of the top-$k$ set at the sampling time. Thus, we define the \gls{k-qaoi} as the average \gls{coud} of the nodes that belong to the top-$k$ set at the sampling time, measured at the deadline $T$. Specifically, the \gls{k-qaoi} is defined as~\footnote{The \gls{k-qaoi} can track both information freshness and the importance of the reading with respect to the top-$k$ set, motivated by the scenario mentioned in Sec.~\ref{sec:scenario}. This metric could be generalized, e.g., by setting the weights depending on the sensor's reading, which is out of scope in this paper.}
\begin{equation}
    \Delta_{k}=\frac{1}{k}\sum_{n \in \Omega_{k}}\Delta_{n}(T),\label{eq:k_QAoI}
\end{equation}
where $\Omega_{k}$ is a subset of nodes whose observed values belong to the top-$k$ set at the sampling time slot $t_{s}$. This definition is equivalent to the average \gls{qaoi} across the top-$k$ nodes. A large \gls{k-qaoi} means the sink holds stale information on top-$k$ set, while a smaller \gls{k-qaoi} means the sink has fresh information on top-$k$ set at the deadline.

To allow each episode to be analyzed in isolation, we assign a fixed age penalty $\Gamma$ to the nodes that fail to transmit their observations before the deadline\footnote{In practice, $\Gamma$ could, for instance, be selected based on the statistics of the penalty value of all nodes, e.g., by averaging the penalty value assigned to each node based on the past data collection history. We argue that the variance of these penalties among top-$k$ nodes is small as the sink might receive data from the nodes even after it exceeds the deadline in the previous episode.}. The value of $\Gamma$ corresponds to the time elapsed since the latest received packet was generated at the deadline when the nodes fail data transmission in the current episode. A small $\Gamma$ means that the sink has partial knowledge on the physical process observed the sensor nodes in the current episode based on the collected data in the previous ones, while a large $\Gamma$ means that the time between consecutive requests is sufficient and the sink has less knowledge on them. 

In practice, having an age that exceeds a certain threshold may not bring any useful information, regardless of the actual age. Thus, we assume the \gls{coud} defined in Eq.~\eqref{eq:AoI_with_CoUD} is upper bounded by a fixed value $A_{\mathrm{max}}$.
Further, as $\Delta_{k}$ defined in Eq.~\eqref{eq:k_QAoI}, is a random variable, in the following, we limit our focus to the expected \gls{k-qaoi}, defined as
\begin{equation}
\ \bar{\Delta}_{k}=\frac{1}{k}\mathbb{E}\left[\sum_{n \in \Omega_{k}}\min(\Delta_{n}(T), A_{\max})\right],\label{eq:bar_k-QAoI}
\end{equation}
where the expectation is taken over the physical process and the transmission of the observations\footnote{The focus of this paper is the \gls{k-qaoi} reduction in a single episode, by applying \gls{cowu}. The minimization for the long-term \gls{k-qaoi} will be kept for future work.}.

\begin{figure}[t]
\centering
\includegraphics[width=0.45\textwidth]{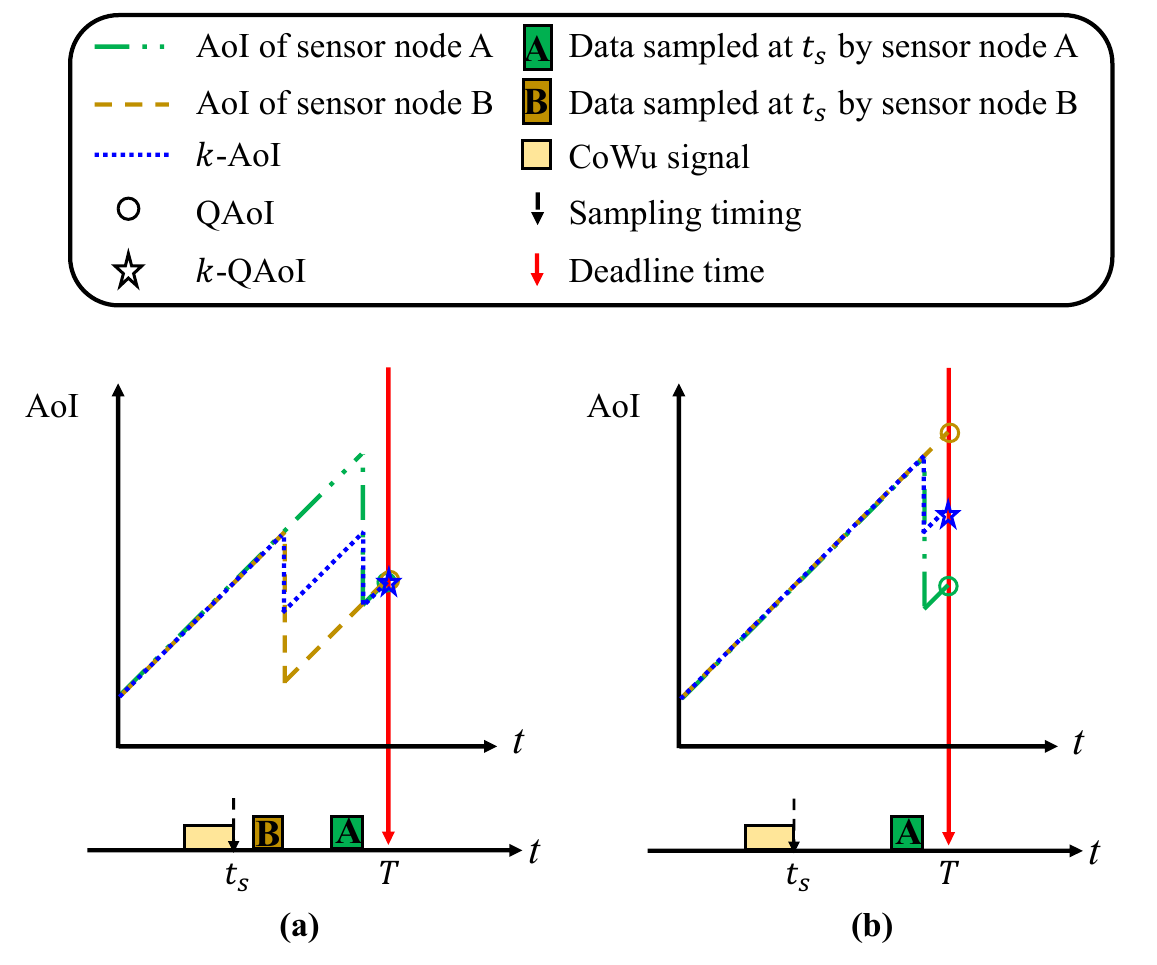}
\caption{An example of the evolution of \gls{aoi}, \gls{qaoi}, and \gls{k-qaoi} with linear age, where $k = 2$ and $\Omega_{k}=\{A, B\}$. (a): The sink receives fresh data from nodes $A$ and $B$ by the deadline. (b): The sink receives data from sensor node $A$ while it fails to collect data from the sensor node $B$.}
\label{Fig:k_QAoI_Evolution}
\end{figure}

An illustration of our introduced \gls{k-qaoi} metrics is shown in Fig.~\ref{Fig:k_QAoI_Evolution} for the linear age case, where the sink performs timely top-$2$ query by applying \gls{cowu} and $\Omega_{k} = \{A, B\}$. In the figure, we show the time evolution of \gls{aoi} for the sensor nodes $A$ and $B$, and the time evolution of $k$-\gls{aoi}, respectively. Here, the $k$-\gls{aoi} can be derived by taking the average value of \gls{aoi} for the nodes belonging to the top-$k$ set at the sampling time, i.e., by taking the average of \gls{aoi} of both sensor nodes A and B in the example of Fig.~\ref{Fig:k_QAoI_Evolution}. In the example of Fig.~\ref{Fig:k_QAoI_Evolution}(a), the sink manages to collect the top-$k$ set by the deadline, which decreases \gls{aoi} of sensor nodes $A$ and $B$ as well as $k$-\gls{aoi} upon its successful data reception at the sink. Thanks to the successful data update by the deadline, the sink can achieve the smaller \gls{qaoi} for both sensor node $A$ and sensor node $B$, leading to the smaller \gls{k-qaoi} in Eq.~\eqref{eq:k_QAoI}. On the other hand, in the example of Fig.~\ref{Fig:k_QAoI_Evolution}(b), the sink fails to collect data from sensor node $B$ by the deadline. In this case, as in the standard definition of \gls{aoi}, the \gls{aoi} of sensor node $B$ monotonically increases, which leads to higher \gls{qaoi}. This failure of data collection by the deadline leads to higher \gls{k-qaoi} than the case of Fig.~\ref{Fig:k_QAoI_Evolution}(a) because the \gls{qaoi} of sensor node $B$ is higher. 

\section{Analysis}\label{sec:Theoretical_Analysis}
In this section, we derive the expected \gls{k-qaoi}, $\bar{\Delta}_{k}$, defined in Eq.~\eqref{eq:bar_k-QAoI} and the total energy consumption of nodes for different wake-up control, defined in Sec.~\ref{sec:energy_model}. 
\subsection{\gls{cowu}}
Let $P_{w}(V_{\mathrm{th}})$ denote the probability that any one node observes a value equal to or more than \gls{cowu} threshold, $V_{\mathrm{th}}$ when the observed value of each node at the sampling time follows a distribution of $g(v;t_{s})$ with a range of [$V_{\mathrm{min}}, V_{\mathrm{max}}$]. We then have 
\begin{equation}
P_{w}(V_{\mathrm{th}})=\int_{V_{\mathrm{th}}}^{V_{\mathrm{max}}}g(v;t_{s})dv,
\end{equation}
and the probability that $w$ out of $N$ nodes wake up follows a binomial distribution:
\begin{equation}
\begin{split}{
\ P_{d}(w)=\binom{N}{w}P_{w}(V_{\mathrm{th}})^{w}(1-P_{w}(V_{\mathrm{th}}))^{N-w}.\label{eq:distribution_nwake}
}\end{split}
\end{equation}

In order to obtain the distribution of the number of successful data transmissions by the deadline under $p$-persistent \gls{csma}, we construct a two-dimensional Markov chain indexed by the time slot. Conditioned on the number of activated nodes $w$, the state space is defined as $\{(w,0), (w, 1), \ldots, (w, L-1), (w-1, 0), \ldots, (1, L-1), (0,0)\}$, where state $(m, l)$ represents the case where $m$ nodes have not completed their transmission, and the transmitting node(s) has been transmitting for $l$ slots. 

We now define the transition probability for the Markov chain. Let $P_{(m, l), (m^{'},l^{'})}$ be the transition probability from the state $(m, l)$ to state $(m^{'},l^{'})$. When the state is $(m, 0)$, $m = 1, 2, \ldots, w$, the channel is idle and $m$ nodes attempt to transmit their packets with probability $p$. If at least one node transmits its packet, the Markov chain transitions to state $(m,1)$. Thus, this transition probability for $m \in \{1, 2, \ldots, w\}$ can be defined as
\begin{equation}
P_{(m, 0), (m, 1)}= 1-(1-p)^m.\label{eq:markov_transmit}
\end{equation}
On the other hand, if none of the nodes transmits, the Markov chain remains in state $(m,0)$, which happens with probability $(1-p)^m$. In states $(m, l)$, $l = 1, \ldots, L-2$, the channel is busy, and $L-l$ slots remain of the current transmission, so the Markov chain transitions to state $(m, l+1)$ with  probability~$1$. %

When the system state is  ($m, L-1$), there are two possible Markov chain transitions. 
The first case is transiting to the state ($m-1, 0$), for which two conditions should be satisfied: (1) only a single node transmits data with probability $p$ while others suppress data with probability $1-p$, under the condition transmission event happens (i.e., at least one node transmits packet), and (2) its transmission succeeds without error. Then, the transition probability $P_{(m, L-1), (m-1, 0)}$, for $m \in \{1, \ldots, w\}$, is given as
\begin{equation}
P_{(m, L-1), (m-1, 0)}=\frac{(1 -e_{c})mp(1-p)^{m-1}}{1-(1-p)^m}.\label{eq:S_n}
\end{equation}
The second case is transiting to the state ($m, 0$), which happens if a collision occurs (i.e., more than two nodes transmit data simultaneously) under the condition that at least one node transmits data or when only one node transmits data, but it experiences a single packet error with probability $e_{c}$. This transmission failure probability is $P_{(m, L-1), (m, 0)}=1-P_{(m, L-1), (m-1, 0)}$, $\forall m \in \{1, \ldots, w\}$. Finally, state $(0, 0)$ is an absorbing state representing the event that all $w$ active users have successfully transmitted their measurement. 

Using the defined Markov chain, we can obtain the distribution of the number of successful transmissions by state evolution from the initial state distribution $\mathbf{\Phi}(0) = (1, 0, 0, \ldots, 0)$ as expressed below:
\begin{equation}
\mathbf{\Phi}(t+1)=\mathbf{\Phi}(t){\mathbf{R}},
\end{equation} 
where $\mathbf{R}$ is $(wL+1) \times (wL+1)$ transition matrix containing the transition probabilities defined above and $\mathbf{\Phi}(t)\in[0,1]^{(wL +1)}$ is the state vector representing the probability of each state in time slot $t$, whose entry corresponding to state $(m, l)$ is denoted as $\mathbf{\phi}_{(m, l)}(t)$. The probability that $w_{s}$ out of the $w$ wake-up nodes succeed by the deadline for a given $\zeta$, denoted as $P_{s}(w_{s}|w, \zeta)$, is then
\begin{equation}\begin{split}
P_{s}(w_{s}|w, \zeta)&=\begin{cases}\phi_{(0,0)}(\zeta),&\text{if}~ w_{s}=w,\\\sum_{l=0}^{L-1}\phi_{(w-w_{s},l)} (\zeta), & \text{otherwise.}\label{eq:n_s_D}
\end{cases}\end{split}\end{equation}

Let us denote the number of nodes belonging to the top-$k$ set at $t_{s}$ that succeed in data transmission by $T$ as $r$. Then, the probability that $r$ nodes ($0 \leq r \leq w_{s} \leq w$) belonging to the top-$k$ set at $t_{s}$ succeed in data transmission by the deadline, given $w$ and $w_{s}$, denoted as $P_{k}(r|w, w_{s})$ can be expressed by considering the possible combinations as follows: 
\begin{equation}\begin{split}
\ &P_{k}(r|w, w_{s})\\&=
\begin{cases}
\delta(r-w_{s}), ~~&\text{if}~ w \leq k,\\
u(w-k-w_{s}+r)\frac{\binom{w_{s}}{r}\binom{w-w_s}{k-r}}{\binom{w}{k}},&\text{if}~ w > k,\label{eq:P_k}
\end{cases}\end{split}\end{equation}
where $\delta(r-w_{s})$ is the Kronecker delta and $u(x)$ is the step function, which outputs 1 if $x \geq 0$ and 0 otherwise.

Using the derived probabilities, we can now compute the expected \gls{k-qaoi}. From the definition of \gls{k-qaoi}, it can be seen that the age of each node at the deadline can only take two values, depending on whether the transmission succeeds or not. Specifically, a successful transmission before the deadline results in an age of $T-u_i(T)=\zeta$, whereas $T-u_i(T)=\Gamma$ if the transmission fails.
Combining this with Eqs.~\eqref{eq:distribution_nwake},~\eqref{eq:n_s_D}, and~\eqref{eq:P_k}, the expected \gls{k-qaoi} of \gls{cowu} for given $\zeta$ and $k$ can be computed as
\begin{equation}
\begin{split}
&\bar{\Delta}_{k}^{\mathrm{Co}}(\zeta) =\sum_{w=0}^{N}P_{d}(w)\sum_{w_{s}=0}^{w}P_{s}(w_{s}|w, \zeta)\sum_{r=0}^{\min(w_{s},k)}P_{k}(r|w, w_{s})\\&\times\frac{r\min(f_s(\zeta),A_{\mathrm{max}})+(k-r)\min(f_{s}(\Gamma), A_{\mathrm{max}})}{k}.\label{eq:k_QAoI_CoWu}
\end{split}
\end{equation}

Next, we analyze the total energy consumption of sensor nodes. We assume that the node activated by a wake-up signal continues the attempt of data transmission until it succeeds even after it exceeds the deadline time following the operation of $p$-persistent \gls{csma}. Thus, the total energy consumption only depends on the value of $V_{\mathrm{th}}$ and transmission probability $p$, but not on the value of $\zeta$. Here, the type of traffic considered in this paper is called \gls{osd} model~\cite{epoch}, in which each node only has a single packet to transmit. 
In~\cite{epoch}, the total energy consumption of nodes operating with $p$-persistent \gls{csma} have been analyzed with the assumption of \gls{osd} and the collision channel, which was extended in~\cite{TGCN_Content}, where the authors also have taken into account the packet errors due to channel impairments with an independent and identical probability of $e_{c}$ for each packet. Let us consider $w$ nodes activated by single wake-up signaling. Then, based on our energy model in Sec.~\ref{sec:energy_model}, $w$ out of $N$ sensor nodes consume energy, while ($N-w$) nodes remain in the sleep state. According to~\cite{TGCN_Content}, the total energy consumption of the sensor nodes with $w$ active nodes, can be expressed as
\begin{equation}
\begin{split}
\ E_{\mathrm{total}}(w|N)&={\xi}_{T}\sum_{m=1}^{w}{\frac{L}{(1-e_{c})(1-p)^{m-1}}}{\delta}\\
&\quad+{\xi}_{R}\sum_{m=1}^{w}{\frac{L-(L-1)(1-p)^{m-1}}{(1-e_{c})p(1-p)^{m-2}}}{\delta},\label{eq:OSD_total_energy}
\end{split}
\end{equation}
where the first sum corresponds to the total energy consumed during the transmission state by $w$ wake-up nodes, and the second sum captures the total energy consumed during the receiving state by $w$ wake-up nodes, as in the energy model described in Sec.~\ref{sec:energy_model}. We refer the reader to~\cite{epoch} for a detailed derivation of Eq.~\eqref{eq:OSD_total_energy}. 
This work assumes $E_{\mathrm{total}} (0) = 0$ as a special case. Since the number of activated nodes $w$ is a random variable, total energy consumption of each node when we employ \gls{cowu} can be expressed as follows:
\begin{equation}
\ E_{\mathrm{total}}^{\mathrm{Co}}(N) = \sum_{w=0}^{N}P_{d}(w)E_{\mathrm{total}}(w|N).
\end{equation}
\subsection{Baseline 1: \gls{rr}}
As a main baseline, we apply \gls{rr}, in which each node transmits its measurement according to a \gls{tdma}-like policy. In this scheme, the sink transmits a single wake-up signal to trigger all nodes equipped with the wake-up receiver at time $t_{\mathrm{sch}}$, which is $NL$ slots prior to the deadline. The nodes receiving the wake-up signal then sample and transmit their measurements in order, i.e., node $j=0, 1, \ldots, N-1$, samples physical process and transmits its measurement at $t_{\mathrm{sch}}+jL$. To derive the \gls{k-qaoi} of \gls{rr}, we first calculate the expected \gls{qaoi} of nodes observing $i \in [1, k]$-th largest values. The transmission time and sampling time of the nodes observing $i$-th largest values is a random variable, uniformly distributed between the duration of [$L$, $NL$] with an interval of $L$. Based on this observation, the expected \gls{qaoi} of $i$-th largest node can be expressed as $\frac{1}{N}\sum_{w=1}^N(1-e_{c})\min(f_{s}(wL), A_{\mathrm{max}})+e_{c}\min(f_{s}(\Gamma), A_{\mathrm{max}})$. Finally, based on the definition of Eq.~\eqref{eq:bar_k-QAoI}, the expected \gls{k-qaoi} of \gls{rr} can be expressed as in Eq.~\eqref{eq:RR_k_QAoI}, in which the value of $k$ is canceled between denominator and numerator:
\begin{equation}\begin{split}
\bar{\Delta}_{k}^{\mathrm{RR}}(N)=&\frac{1}{N}\sum_{w=1}^N(1-e_{c})\min(f_{s}(wL), A_{\mathrm{max}})\\&+e_{c}\min(f_{s}(\Gamma), A_{\mathrm{max}}).\label{eq:RR_k_QAoI}\end{split}
\end{equation}
The total energy consumption of \gls{rr} can be easily derived based on Eq.~\eqref{eq:Energy model} and expressed as
\begin{equation}
E_{\mathrm{total}}^{\mathrm{RR}} (N) = \xi_{T}NL\delta.\label{eq:RR_ene}
\end{equation}
 
\subsection{Baseline 2: Probability $q$-based Random Wake-up}
As a second baseline, we consider the probability $q$-based random wake-up, which we call $q$-Wu. In this scheme, the sink transmits a wake-up signal $\zeta$ slots prior to the deadline. Each sensor node receiving the wake-up signal randomly decides with probability $q$ whether it wakes up (regardless of its content), and otherwise stays silent. The node woken up transmits its data following the $p$-persistent \gls{csma}. The probability that $\nu$ out of $N$ nodes wake up can be expressed as $P_{x}(\nu) = \binom{N}{\nu}q^{\nu}(1-q)^{N-\nu}$. By applying Eq.~\eqref{eq:n_s_D}, the probability $\nu_{s}$ out of $\nu$ nodes succeeds in data transmission by the deadline $T$, given $\nu$ and $\zeta$, can be described as $P_{s}(\nu_{s}|\nu, \zeta)$. Finally, the probability that $c$ out of $\nu_{s}$ nodes belonging to the top-$k$ set at $t_{s}$ successfully transmit can be expressed as
\begin{equation}
\ P_{y}(c|\nu, \nu_{s}) =\frac{\binom{k}{c}\binom{N-k}{\nu_{s} - c}}{\binom{N}{\nu_{s}}}.
\end{equation}
Finally, the expected \gls{k-qaoi} for $q$-Wu can be described as:
\begin{equation}
\begin{split}
\bar{\Delta}_{k}^{\mathrm{q}}(\zeta) =\sum_{\nu=0}^{N}P_{x}(\nu)\sum_{\nu_{s}=0}^{\nu}P_{s}(\nu_{s}|\nu, \zeta)\sum_{c=0}^{\min(\nu_{s},k)}P_{y}(c|N, \nu_{s})\\\times\frac{c\min(f_s(\zeta),A_{\mathrm{max}})+(k-c)\min(f_{s}(\Gamma), A_{\mathrm{max}})}{k}.\label{eq:k_QAoI_qWu}
\end{split}
\end{equation}
The total energy consumption for $q$-Wu can be expressed as:
\begin{equation}
\ E_{\mathrm{total}}^{\mathrm{q}}(N) = \sum_{\nu=0}^{N}P_{x}(\nu)E_{\mathrm{total}}(\nu|N).
\end{equation}
\subsection{Baseline 3: Genie-aided Scheme}
In order to obtain a lower bound on total energy consumption and \gls{k-qaoi}, we consider as a final baseline a Genie-aided scheme, in which we assume the sink perfectly knows the observations of the nodes belonging to the top-$k$ set at $t_{s}$, $\Omega_{k}$. In this scheme, the sink transmits wake-up signal to the nodes belonging to the top-$k$ set $\Omega_{k}$, to notify the timing of sampling and its activation. Specifically, it transmits wake-up signals, including \glspl{id} in $\Omega_{k}$, one by one. The total number of signaling transmissions is $k$. The wake-up receiver detecting the wake-up signal, including its own \gls{id}, regards itself as belonging to the top-$k$ set and records its own order, say $l~\in~[1, k]$. Then, after receiving the wake-up signal from the sink at $t_{G} = T- kL$, the node whose order $l$ samples and transmits its data at $t_{G} +(l-1)L$. Then, the \gls{k-qaoi} of the Genie-aided scheme is expressed as: 
\begin{equation}\begin{split}
\tilde{\Delta}_{k}(k) &= \frac{1}{k}\sum_{w = 1}^{k}(1-e_{c})\min(f_{s}(wL), A_{\mathrm{max}}) \\&+e_c\min(f_{s}(\Gamma), A_{\mathrm{max}}).\label{eq:genie_k_QAoI}
\end{split}\end{equation}
Let $g(\Omega_{k}^{*})$ be the \gls{k-qaoi} when the sink collects data from the subset of nodes belonging to the top-$k$ set $\Omega_{k}^{*}$, where $\Omega_{k}^{*} \subseteq \Omega_{k}$. Then, the following statement is always true: $g(\Omega_{k}^{*}) \geq g(\Omega_{k})$. Under the collision model, at most one node can transmit its packet with the length of $L$. Then, the minimum value of \gls{k-qaoi} that the system can achieve is expressed as Eq.~\eqref{eq:genie_k_QAoI}, in which only the $k$ nodes belonging to the top-$k$ set access the channel in order and deliver its fresh data with respect to the deadline. Therefore, Eq.~\eqref{eq:genie_k_QAoI} is the lower bound of \gls{k-qaoi}. 
The total energy consumption of the Genie-aided scheme can be expressed as: 
\begin{equation}
\tilde{E}_{\mathrm{total}}(k) = \xi_{T}kL\delta.\label{eq:genie_ene}
\end{equation}

Accordingly, the lower bound on the energy consumption can be derived as follows. 
Here, we are interested in a lower bound on the energy consumption given $k$ that achieves the lower bound of \gls{k-qaoi} $\tilde{\Delta}_{k}(k)$ in Eq.~\eqref{eq:genie_k_QAoI}. 
Let $O(K_{1})$ be the total energy consumption when the sink collects data from $K_{1} \geq k$ sensor nodes under the same access protocol. Then, the following statement is always true: $O(K_{1}) \geq O(k)$. Therefore, Eq.~\eqref{eq:genie_ene} is the lower bound of total energy consumption under the lower bound of \gls{k-qaoi}, in which the nodes in $\Omega_{k}$ activate their main radio only in their allocated slots.

Finally, it is worth emphasizing that the genie-aided scheme cannot be implemented in practice, since it relies on the assumption of having a perfect and instantaneous observation of the status of the nodes, including those that belong to the top-$k$ set.

\section{Optimization}\label{sec:Optimization}
\subsection{Problem Formulation}\label{sec:problem_Formulation}
We are interested in how much energy consumption can be reduced by applying \gls{cowu}, while guaranteeing a certain level of \gls{k-qaoi}. As the performance of \gls{cowu} depends on the value of $V_{\mathrm{th}}$, transmission probability of $p$,  and the timing of \gls{cowu} signaling $\zeta$, it is desirable to optimize these parameters in terms of \gls{k-qaoi} and total energy consumption. This can be formulated as the energy minimization problem under the \gls{k-qaoi} constraint, as expressed below: 
\begin{IEEEeqnarray}{rCl}
 &\min_{V_{\mathrm{th}},p,\zeta} &{E_{\mathrm{total}}^{\mathrm{Co}}(N) }\label{eq:promlem_formulation1} \\ 
 &\text{s.t.}&\bar{\Delta}_{k}^{\mathrm{Co}}(\zeta)~\leq~\gamma_{\mathrm{th}},\\
 & &V_\mathrm{th}\in[V_{\mathrm{min}}, V_{\mathrm{max}}]\\
 & &p~\in~(0, 1]\\
 &&\zeta~\in~\mathbb{Z}^+\label{eq:promlem_formulation_last}
\end{IEEEeqnarray}
As obtaining the exact optimal values is computationally and analytically difficult, we introduce an approximate solution, as we will describe in Sec.~\ref{sec:Approximation}, in which we obtain the set of optimal parameters with two steps based on an analysis and a grid search.

\subsection{Approximate Solution}\label{sec:Approximation}
In order to tackle the problem mentioned in Sec.~\ref{sec:problem_Formulation}, we introduce an approximate solution to derive the optimal sets of parameter $\{V_{\mathrm{th}}, p, \zeta\}$ for Eqs.~\eqref{eq:promlem_formulation1}--\eqref{eq:promlem_formulation_last}. Specifically, we first obtain the optimal transmission probability $p$ analytically (c.f. Sec.~\ref{sec:step_1_optimal_transmission}) and then using the optimal transmission probability, we obtain the optimal parameter of $\{\zeta, V_{\mathrm{th}}\}$ with grid-search (c.f. Sec.~\ref{sec:step_2_optimall_zeta_and_Vth}).
\subsubsection{Step 1. Optimizing the Transmission Probability $p$}\label{sec:step_1_optimal_transmission}
Here, we first optimize the parameter of $p$ in terms of data collection delay. This is because a smaller delay is important to realize information freshness at the deadline, by which the sink can activate nodes at a much later time to obtain smaller \gls{k-qaoi}. 

In the \gls{osd} model, the number of active users gradually decreases as packets are successfully transmitted. In~\cite{epoch}, in order to analyze the delay and energy consumption of the nodes operating with $p$-persistent \gls{csma} protocol under \gls{osd}, the authors introduced the notion of an ``epoch,'' which is defined as the time required for a single sensor node succeeding in data transmission when the $m$ active nodes compete for the channel. Then, the authors analyzed the expected time for an \emph{epoch} with $m$ active nodes analytically, denoted as $\mathbb{E}[T_m]$, which can be expressed as~\cite{epoch,TGCN_Content}:
\begin{equation}
    \mathbb{E}[T_m] = \frac{L-(L-1)(1-p)^m}{(1-e_{c})mp(1-p)^{m-1}}{\delta}.\label{eq:epoch_delay}
\end{equation}
Then, data collection delay when we employ $p$-persistent \gls{csma} in \gls{osd} can be expressed as follows~\cite{epoch,TGCN_Content}:
\begin{equation}T_{d} (w)
=\sum_{m=1}^{w}\mathbb{E}[T_m],
\label{eq:OSDdelay}
\end{equation}
where this work assumes $T_{d} (0)=0$, as a special case. 

From Eq.~\eqref{eq:OSDdelay}, we can observe that, to minimize the data collection delay $T_{d} (w)$ in Eq.~\eqref{eq:OSDdelay}, we must minimize the delay $\mathbb{E}[T_m]$ in Eq.~\eqref{eq:epoch_delay}. As $\mathbb{E}[T_m]$ in Eq.~\eqref{eq:epoch_delay} depends on the value of $p$ and $m$, we take an approach to optimize $p$ in $\mathbb{E}[T_m]$ for each $m$ step-by-step so that $\mathbb{E}[T_m]$ can be minimized. 
Let us denote the optimal transmission probability $p$, which minimizes $\mathbb{E}[T_m]$ as $p_{\mathrm{opt}}^{*}(m)$. Then, we can formulate the optimization problem for each $m \in \{1, 2, \ldots,w\}$ as follows: 
\begin{equation}
 p_{\mathrm{opt}}^{*}(m) = \argmin_{p \in (0, 1]}  \mathbb{E}[T_m].\label{eq:dynamic_P_formulation}
\end{equation}
We analytically solve the problem in Eq.~\eqref{eq:dynamic_P_formulation} to obtain optimal transmission probability. Specifically, we set the first order derivative of $\mathbb{E}[T_m]$ with respect to $p$ to be 0, i.e., $\frac{\mathrm{d}\mathbb{E}[T_m]}{\mathrm{dp}} = 0$, for the case of $L > 1$. After several steps of the calculation, we can obtain the following results: $(1-p)^m = \frac{L}{L-1}(1-mp)$. Then, by applying the second order approximation of Taylor expansion around $p = 0$, we can obtain the following approximation: $(1-p)^m \approx 1 -mp +\frac{m(m-1)}{2}p^2$. Finally, we can obtain the approximate optimal transmission probability for the case of the active number of wake-up nodes $m > 1$, as in~\cite{epoch}, as $\frac{\sqrt{m^2 + 2m(m-1)(L-1)}-m}{m(m-1)(L-1)}$. When $m =1$, with Eq.~\eqref{eq:OSDdelay}, data collection delay can be expressed as $\mathrm{E}[T_1]=\frac{L-1}{1-e_c}\delta + \frac{1}{p(1-e_c)}\delta$. Clearly, this is a decreasing function for $p \in (0, 1]$, which can be minimized when $p =1$. By combining these results, we can express the optimal transmission probability for each $m$ as follows: 
\begin{equation}
p_{\mathrm{opt}}^{*}(m) = \begin{cases}  1, & m = 1,  \\\approx \frac{\sqrt{m^2 + 2m(m-1)(L-1)}-m}{m(m-1)(L-1)},& m > 1.\label{eq:dynamic_p}
\end{cases}\end{equation}
The \gls{k-qaoi} and total energy consumption of \gls{cowu} applying dynamic transmission probability $p_{\mathrm{opt}}^{*}(m)$ can be calculated by substituting $p_{\mathrm{opt}}^{*}(m)$ into Eqs.~\eqref{eq:markov_transmit} and \eqref{eq:S_n}, and into the Eq.~\eqref{eq:OSD_total_energy} for each $m$. 

Note that, in order to apply the optimal transmission probability $p_{\mathrm{opt}}^{*}(m)$ in Eq.~\eqref{eq:dynamic_p}, each node must know the number of active nodes given time. In practice, this can be realized through the reception of broadcasted \gls{ack}. Specifically, given the number of wake-up nodes $w$, each node updates the number of active nodes and its transmission probability upon the reception of \gls{ack}.

\subsubsection{Step 2. Optimizing the parameters $\zeta$ and $V_{\mathrm{th}}$ for \gls{cowu} signal}\label{sec:step_2_optimall_zeta_and_Vth}
 We then obtain the optimal set of parameter $\mathbb{C}(N)= \{\zeta, V_{\mathrm{th}}\}$ when we apply  $p_{\mathrm{opt}}^{*}$ shown in Eq.~\eqref{eq:dynamic_p} based on grid search, which are required to satisfy 
\begin{equation}\begin{split}
\ \mathbb{C}(N)=\min_{\{\zeta, V_{\mathrm{th}}\}}{E_{\mathrm{total}}^{\mathrm{Co}}(N) } &\\\text{s.t.}~\bar{\Delta}_{k}^{\mathrm{Co}}(\zeta)~\leq~\gamma_{\mathrm{th}},\label{eq:min_set_ene}
\end{split}\end{equation} 
where we select the value of $V_{\mathrm{th}}$ out of the range from 0 to 50 with a step of 0.5, and that of $\zeta$ out of the range from 10 to 1000 with a step of 10. Note that we rely on grid-search methods because the problem in Eq.~\eqref{eq:min_set_ene} is non-convex and cannot be solved analytically.

\subsection{Maximum $k$}
We are further interested in the scenario and region, in which applying \gls{cowu} is beneficial in terms of total energy consumption and \gls{k-qaoi}. Specifically, we investigate the maximum value of $k$, under which \gls{cowu} can satisfy the both the constraint of \gls{k-qaoi} $\gamma_{\mathrm{th}}$ and that of total energy consumption, denoted as $\epsilon_{\mathrm{th}}$, which can be formalized as: 
\begin{IEEEeqnarray}{rCl}
k_\text{max}^*:\quad & \max{} & k\label{eq:maximum_k_1}\\
&\text{subject to}\quad & E_{\mathrm{total}}^{\mathrm{Co}}(N)~\leq~\epsilon_{\mathrm{th}}\\
&&\bar{\Delta}_{k}^{\mathrm{Co}}(\zeta)~\leq~\gamma_{\mathrm{th}}.\label{eq:maximum_k_3}
\end{IEEEeqnarray}
Because Eqs.~\eqref{eq:maximum_k_1}--\eqref{eq:maximum_k_3} are non-convex,
we resort to the approximate solution as in Sec.~\ref{sec:Approximation} to obtain the maximum value of $k$, in which results of average \gls{k-qaoi} and total energy consumption are evaluated for a wide range of parameters $\{\zeta, V_{\mathrm{th}}\}$. Here, we apply the optimal dynamic transmission probability $p_{\mathrm{opt}}^{*}$ obtained in Eq.~\eqref{eq:dynamic_p}, and select the value of $V_{\mathrm{th}}$ out of the range from 0 to 50 with a step of 2, and that of $\zeta$ out of the range from 50 to 500 with a step of 50 for the different number of nodes $N$. 
\section{Numerical Results}\label{sec:sim}
This section investigates the performance of \gls{cowu} for different parameters, including $\zeta$, $V_{\mathrm{th}}$, $k$, $\alpha$, and $\Gamma$, presenting numerical results obtained by both theoretical analysis and simulations. The parameters used for numerical evaluations are shown in Table~\ref{table:para}. The procedures of wake-up/data collections follow the system model given in Sec.~\ref{sec:sys}, including the actual operations of $p$-persistent \gls{csma} protocol. The \gls{k-qaoi} and total energy consumption of sensor nodes are recorded when the sink completes the top-$k$ query. For simplicity of analysis, we assume the physical process at the sampling time $t_{s}$ $g(v;t_{s})$ follows uniform distribution, i.e., $g(v;t_{s}) = \frac{1}{V_{\mathrm{max}}-V_{\mathrm{min}}}$, for $v\in[V_\mathrm{min}, V_\mathrm{max}]$, $0$ otherwise. Numerical results obtained by Monte Carlo simulation illustrate the average values obtained after repeating the sampling, query, and transmission processes over $10^{4}$ rounds.

\begin{table}[!tb]
\caption{Parameters used for Numerical Evaluations.}
\begin{center}
\begin{tabular}{|c|c| p{10zw}} \hline
Parameters & Values  \\ \hline \hline Data transmission rate &
100~kbps \\ \hline Length of packet in time slots~$L$ & 10 \\
\hline Time slot length~$\delta$ & 320~${\mu}$sec \\ \hline Power
consumption in Transmit state~$\xi_{T}$ &
55~mW~\cite{tamura2019low} \\ \hline Power consumption in Receive
state~$\xi_{R}$ & 50~mW~\cite{tamura2019low} \\ \hline
Distribution of observed value~$[V_{\mathrm{min}},V_{\mathrm{max}}]$ & $[0,50]$ \\
\hline
Maximum value of age $A_{\max}$& 5000\\
\hline
\end{tabular}
\label{table:para}
\end{center}
\end{table}

\subsection{Achievable Region for Different Wake-up Scheme}
\begin{figure}[t]
\centering
%
%
\definecolor{mycolor1}{rgb}{0.74902,0.00000,0.74902}%
\definecolor{mycolor2}{rgb}{0,0.15,0.74}%
\begin{tikzpicture}

\begin{axis}[%
width=7cm,
height=4cm,
scale only axis,
xmin=0,
xmax=64,
xtick={0, 10, 20, 30, 40, 50, 60},
scaled ticks=false,
xlabel style={font=\color{white!15!black}},
xlabel={Total energy consumption [mJ]},
ymin=0,
ymax=1200,
ylabel style={font=\color{white!15!black}},
ylabel={k-QAoI},
axis background/.style={fill=white},
legend style={at={(-0.1,1)}, anchor=south west, legend cell align=left, align=left, draw=none, legend columns=2}
]

\addplot [color=orange,  line width=1.5pt]
  table[row sep=crcr]{%
62.5623033972654	410.390127925526\\
38.0839069142048	333.75842102372\\
21.0688087988001	305.617180810864\\
9.8627246359662	319.886067407228\\
3.17514868517466	495.232881756172\\
0	1000\\
};
\addlegendentry{\fontsize{8},\text{CoWu (Theory, $\zeta = 250$)}}

\addplot [color=orange, line width=1.5pt, only marks, mark size=1.5pt, mark=diamond, mark options={solid, orange}]
  table[row sep=crcr]{%
63.0816432000002	411.944\\
38.5082992000001	332.858\\
21.2181087999999	306.02\\
9.83550560000003	319.46\\
3.18920319999999	493.438\\
0	1000\\
};
\addlegendentry{\fontsize{8},CoWu (Simulation, $\zeta = 250$)}



\addplot [color=cyan, dotted, line width=1.5pt]
  table[row sep=crcr]{%
62.5623033972651	895.330059863164\\
38.0839069142049	896.829913971122\\
21.0688087988001	913.240202627209\\
9.86272463596615	940.190899246392\\
3.17514868517468	970.001253636225\\
0	1000\\
};
\addlegendentry{\fontsize{8},$\text{$q$-Wu (Theory, $\zeta = 250$)}$}

\addplot [color=cyan, line width=1.5pt, only marks, mark size=1.5pt, mark=x, mark options={solid, cyan}]
  table[row sep=crcr]{%
62.8269168000002	897.01\\
37.8825968	898.645\\
21.2033743999999	913.72\\
9.93871199999994	938.065\\
3.15877279999999	970.825\\
0	1000\\
};
\addlegendentry{\fontsize{8},$q$-Wu (Simulation, $\zeta = 250$)}

\addplot [color=black,only marks, mark size=1.5pt, mark=star, mark options={solid, rotate=90, black}]
  table[row sep=crcr]{%
17.6000000000000	505\\
};
\addlegendentry{\fontsize{8},$\text{Round-robin scheduling}$}

\addplot [color=black,only marks, mark size=1.5pt, mark=pentagon*, mark options={solid, rotate=90, magenta}]
  table[row sep=crcr]{
  0.880000000000000 30\\
  };
\addlegendentry{\fontsize{8},$\text{Genie-aided Scheme}$}
\end{axis}
\end{tikzpicture}%
\caption{The achievable set of total energy consumption and \gls{k-qaoi} for the different data collection methods (Linear age).}
\label{Fig:ene_k_QAoI_graph}
\end{figure}
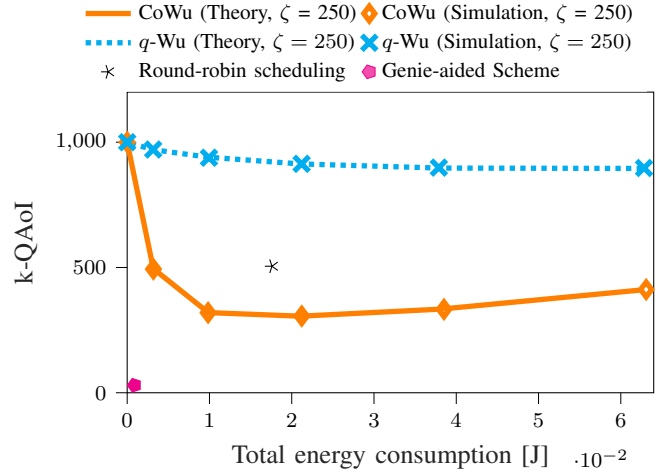

Fig.~\ref{Fig:ene_k_QAoI_graph} shows the achievable set of total energy consumption and \gls{k-qaoi} of \gls{cowu}, \gls{rr}, $q$-Wu, and Genie-aided scheme, where we set $N = 100$, $k = 5$, $e_{c} = 0$, $p = 0.0606$, $\Gamma =1000$, and $\zeta = 250$ for the linear age case, in Eq.~\eqref{eq:cost_function}. Here, the value of $p = 0.0606$ corresponds to the back-off window size of 32 \cite{epoch, TGCN_Content}. In this evaluation, we vary the value of $V_{\mathrm{th}}$ for \gls{cowu} and the parameter of $q$ for $q$-Wu, and plot the achievable set for each result. From this figure, we can first see that the results for \gls{cowu} and $q$-Wu obtained with theoretical analysis coincide with simulation results, which validates our analysis. Next, from Fig.~\ref{Fig:ene_k_QAoI_graph}, we can see that the \gls{k-qaoi} of \gls{cowu} can be reduced at the cost of increasing energy consumption of sensor nodes, i.e., by setting the threshold of \gls{cowu} to a lower value. This is because the top-$k$ nodes at the sampling time are more likely to be woken up by this wake-up signal. However, too small $V_{\mathrm{th}}$ increases the \gls{k-qaoi} because it causes many nodes simultaneously to wake up, by which data collection is more likely to fail by the deadline. The same characteristic can be observed for $q$-Wu, in which the \gls{k-qaoi} is slightly improved by setting the high wake-up probability $q$. However, it is worth emphasizing that the performance of $q$-Wu is always worse than that of \gls{cowu}, as we can see in Fig.~\ref{Fig:ene_k_QAoI_graph}. In addition, from Fig.~\ref{Fig:ene_k_QAoI_graph}, we can observe that the Genie-aid can always achieve better performance than any other scheme, thanks to the exploitation of perfect knowledge of the top-$k$ nodes at the sampling time. Further, from Fig.~\ref{Fig:ene_k_QAoI_graph}, we can see the gap between the \gls{cowu} and the genie-aided scheme, i.e., from the ideal cases, which will bring further motivations to design an efficient communication protocol, which will be kept for future work. Finally, we can see that \gls{cowu} can achieve smaller \gls{k-qaoi} than \gls{rr} while consuming less amount of energy, as long as we can select the appropriate threshold of $V_{\mathrm{th}}$. As mentioned above, $q$-Wu cannot achieve higher performance than \gls{cowu}; hereafter, we compare the performance of \gls{cowu} with that of \gls{rr}.

\subsection{Impact of Different Parameters on the Performance of \gls{cowu}}
\subsubsection{Impact of Timing of \gls{cowu} Signal $\zeta$}\label{sec:timing_of_zeta}
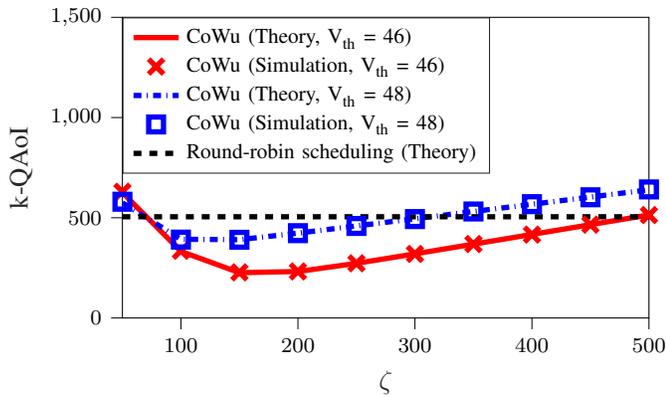
\begin{figure}[t]
\centering
%
%
\begin{tikzpicture}

\begin{axis}[%
width=7cm,
height=4cm,
scale only axis,
xmin=50,
xmax=500,
xlabel style={font=\color{white!15!black}},
xlabel={$\zeta$},
ymin=0,
ymax=1500,
ylabel style={font=\color{white!15!black}},
ylabel={k-QAoI},
axis background/.style={fill=white},
legend style={at={(0,1)},anchor=north west,legend cell align=left, align=left, draw=white!15!black}
]
\addplot [color=red, line width=1.5pt]
  table[row sep=crcr]{%
50	625.47742707595\\
100	332.283083220004\\
150	225.847166757901\\
200	231.839052546167\\
250	272.455358128527\\
300	319.886067407228\\
350	368.332318600538\\
400	416.906924031384\\
450	465.49634431109\\
500	514.087404200214\\
};
\addlegendentry{\fontsize{8},CoWu (Theory, $V_{\textrm{th}} = 46$)}

\addplot [color=red, line width=1.5pt, only marks, mark size=1.5pt, mark=x, mark options={solid, red}]
  table[row sep=crcr]{%
50	629.785\\
100	335.35\\
150	227.707\\
200	232.176\\
250	272.845\\
300	319.46\\
350	368.837\\
400	415.996\\
450	465.631\\
500	513.48\\
};
\addlegendentry{\fontsize{8},CoWu (Simulation, $V_{\mathrm{th}} = 46$)}

\addplot [color=blue, dashdotted, line width=2.0pt]
  table[row sep=crcr]{%
50	574.445741200155\\
100	392.633954356738\\
150	390.803253798322\\
200	423.364829104063\\
250	459.190598536137\\
300	495.232881756172\\
350	531.287097384237\\
400	567.341908877729\\
450	603.396748566236\\
500	639.451589550099\\
};
\addlegendentry{\fontsize{8},CoWu (Theory, $V_{\mathrm{th}} = 48$)}

\addplot [color=blue, line width=1.5pt, only marks, mark size=1.5pt, mark=square, mark options={solid, blue}]
  table[row sep=crcr]{%
50	580.005\\
100	391.042\\
150	390.142\\
200	423.168\\
250	459.205\\
300	493.438\\
350	531.324\\
400	567.904\\
450	602.196\\
500	641.38\\
};
\addlegendentry{\fontsize{8},CoWu (Simulation, $V_{\mathrm{th}} = 48)$}

\addplot [color=black, dashed, line width=1.5pt]
  table[row sep=crcr]{%
50	505\\
100	505\\
150	505\\
200	505\\
250	505\\
300	505\\
350	505\\
400	505\\
450	505\\
500	505\\
};
\addlegendentry{\fontsize{8},$\text{RR-scheduling (Theory)}$}

\end{axis}

\end{tikzpicture}%
\caption{\gls{k-qaoi} of \gls{cowu} against $\zeta$, where $V_{\mathrm{th}} = 46$ and $V_{\mathrm{th}} = 48$ (Linear age).}
\label{Fig:k-QAoI_against_zeta}
\end{figure}

\begin{figure}[t]
\centering
%
%
\begin{tikzpicture}

\begin{axis}[%
width=7cm,
height=4cm,
scale only axis,
xmin=50,
xmax=500,
xlabel style={font=\color{white!15!black}},
xlabel={$\zeta$},
ymin=0,
ymax=8000,
ylabel style={font=\color{white!15!black}},
ylabel={k-QAoI},
axis background/.style={fill=white},
legend style={at={(0,1)},anchor=north west,legend cell align=left, align=left, draw=white!15!black}
]
\addplot [color=red, line width=1.5pt]
  table[row sep=crcr]{%
50	3029.50596916963\\
100	1295.20166326614\\
150	463.5423014971\\
200	250.459085552578\\
250	292.701926252077\\
300	533.039666238638\\
350	1205.75023054086\\
400	3036.87914846389\\
450	5000.00000000002\\
500	5000.00000000002\\
};
\addlegendentry{\fontsize{8},CoWu (Theory, $V_{\mathrm{th}} = 46$)}

\addplot [color=red, line width=1.5pt, only marks, mark size=1.5pt, mark=x, mark options={solid, red}]
  table[row sep=crcr]{%
50	3060.46676208098\\
100	1327.39889519855\\
150	462.5861607156\\
200	252.047792253799\\
250	290.758574382688\\
300	537.229581267638\\
350	1199.64549108808\\
400	3033.6507037454\\
450	5000\\
500	5000\\
};
\addlegendentry{\fontsize{8},CoWu (Simulation, $V_{\mathrm{th}} = 46$)}

\addplot [color=blue, dashdotted, line width=1.5pt]
  table[row sep=crcr]{%
50	2761.01045068495\\
100	1630.0558528021\\
150	1430.16837174674\\
200	1434.66340490548\\
250	1500.9005533638\\
300	1684.71033024362\\
350	2184.57366893887\\
400	3543.35413114447\\
450	4999.99999999998\\
500	4999.99999999998\\
};
\addlegendentry{\fontsize{8},CoWu (Theory, $V_{\mathrm{th}} = 48$)}

\addplot [color=blue, line width=1.5pt, only marks, mark size=1.5pt, mark=square, mark options={solid, blue}]
  table[row sep=crcr]{%
50	2798.25690314563\\
100	1623.12053529626\\
150	1408.26258067513\\
200	1407.72512048002\\
250	1493.42369703056\\
300	1674.57674633346\\
350	2183.78019717435\\
400	3540.88325319936\\
450	5000\\
500	5000\\
};
\addlegendentry{\fontsize{8},CoWu (Simulation, $V_{\mathrm{th}} = 48$)}

\addplot [color=black, dashed, line width=1.5pt]
  table[row sep=crcr]{%
50	3144.85418871071\\
100	3144.85418871071\\
150	3144.85418871071\\
200	3144.85418871071\\
250	3144.85418871071\\
300	3144.85418871071\\
350	3144.85418871071\\
400	3144.85418871071\\
450	3144.85418871071\\
500	3144.85418871071\\
};
\addlegendentry{\fontsize{8},$\text{RR-scheduling (Theory)}$}

\end{axis}

\end{tikzpicture}%
\caption{\gls{k-qaoi} of \gls{cowu} against $\zeta$, where $V_{\mathrm{th}} = 46$ and $V_{\mathrm{th}} = 48$ (Exponential age with $\alpha = 0.02$).}
\label{Fig:k-QAoI_against_zeta_EXP}
\end{figure}
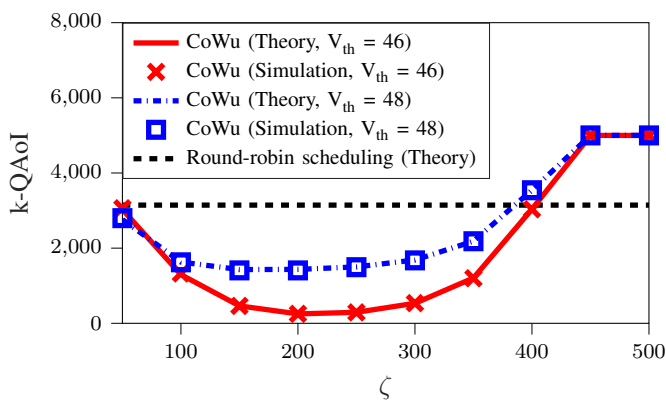

First, we investigate the impact of the timing of \gls{cowu} signal in terms of \gls{k-qaoi}. Figs.~\ref{Fig:k-QAoI_against_zeta} and~\ref{Fig:k-QAoI_against_zeta_EXP} show \gls{k-qaoi} of \gls{cowu} against $\zeta$ for the linear age function and for the exponential age function defined in Eq.~\eqref{eq:cost_function} with $\alpha = 0.02$, respectively, where we set $N = 100$, $k = 5$, $e_{c} = 0$, $p = 0.0606$, $\Gamma =1000$, and $V_{\mathrm{th}}~=~46,~48$. 
From these figures, we can first see that the results for \gls{cowu} obtained with theoretical analysis coincide with simulation results for both the linear age case and exponential age case, which validates our analysis. 
Next, from Figs.~\ref{Fig:k-QAoI_against_zeta} and~\ref{Fig:k-QAoI_against_zeta_EXP}, we can also see that there is an optimal value of $\zeta$ in terms of \gls{k-qaoi} when the value of $V_{\mathrm{th}}$ is fixed. Let us denote the optimal value of $\zeta$ as $\zeta_{\mathrm{opt}}$. For $\zeta < \zeta_{\mathrm{opt}}$, we can see that the \gls{k-qaoi} becomes larger as $\zeta$ decreases, because the number of nodes that fail their data transmission by the deadline $T$ increases due to the congestion. For $\zeta > \zeta_{\mathrm{opt}}$, most users complete their transmission, but the sensed values become obsolete at the deadline, also leading to an increase of \gls{k-qaoi}. This result illustrates the importance of the timing of the wake-up signal so as to minimize the \gls{k-qaoi} at the deadline. Further, from Fig.~\ref{Fig:k-QAoI_against_zeta_EXP},  we can see that the \gls{k-qaoi} of the exponential case is much larger than that of the linear age case shown in Fig.~\ref{Fig:k-QAoI_against_zeta} and it reaches maximum age $A_{\max}$ when the value of $\zeta$ is equal to or larger than 450. This is because the collected data in the exponential age case becomes obsolete much faster than the linear age case (c.f.~Sec.~\ref{Sec:timing_for_alpha_exp}). Finally, we can clearly see \gls{cowu} achieves smaller \gls{k-qaoi} than \gls{rr} by transmitting a wake-up signal at the right timing against the deadline, i.e., by selecting the appropriate parameter of $\zeta$.

\subsubsection{Timing Analysis Considering Different Characteristics of Physical Process}\label{Sec:timing_for_alpha_exp}
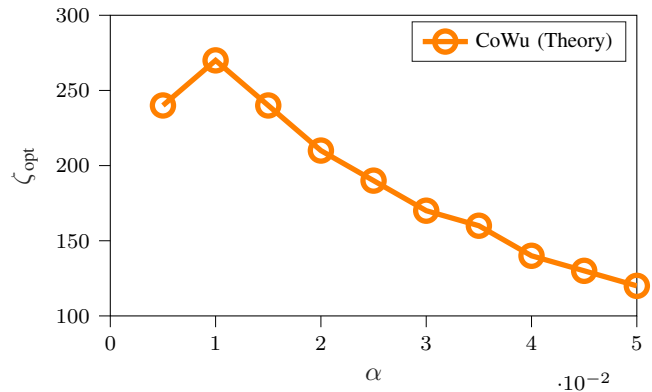
\begin{figure}[t]
\centering
%
%
\begin{tikzpicture}

\begin{axis}[%
width=7cm,
height=4cm,
scale only axis,
xmin=0.005,
xmax=0.05,
xlabel style={font=\color{white!15!black}},
xlabel={$\alpha$},
xtick={0.01, 0.02, 0.03, 0.04, 0.05},
xticklabels={
  \(\displaystyle {0.01}\),
  \(\displaystyle {0.02}\),
  \(\displaystyle {0.03}\),
  \(\displaystyle {0.04}\),
  \(\displaystyle {0.05}\), 
},
scaled ticks=false,
ymin=100,
ymax=300,
ylabel style={font=\color{white!15!black}},
ylabel={$\zeta{}_{\text{opt}}$},
axis background/.style={fill=white},
legend style={legend cell align=left, align=left, draw=white!15!black}
]
\addplot [color=orange, line width=1.5pt, mark size=1.5pt, mark=o, mark options={solid, orange}]
  table[row sep=crcr]{%
0.005	240\\
0.01	270\\
0.015	240\\
0.02	210\\
0.025	190\\
0.03	170\\
0.035	160\\
0.04	140\\
0.045	130\\
0.05	120\\
};
\addlegendentry{CoWu (Theory)}

\end{axis}

\end{tikzpicture}%
\caption{Optimal timing of \gls{cowu} signal $\zeta_{\mathrm{opt}}$ against $\alpha$ (Exponential age with $\alpha = 0.02$).}
\label{Fig:zeta_opt_Against_alpha}
\end{figure}

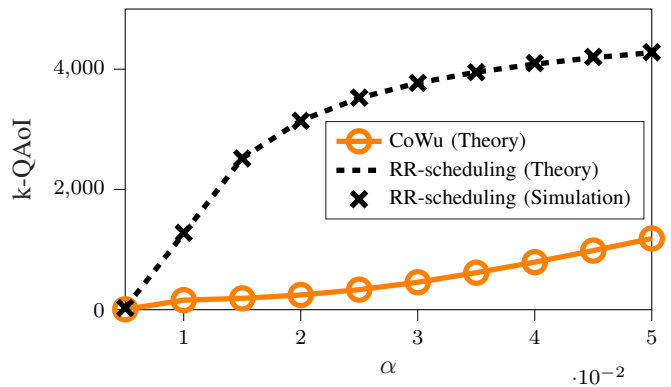
\begin{figure}[t]
\centering
%
%
\begin{tikzpicture}

\begin{axis}[%
width=7cm,
height=4cm,
scale only axis,
xmin=0.005,
xmax=0.05,
xlabel style={font=\color{white!15!black}},
xlabel={$\alpha$},
xtick={0.01, 0.02, 0.03, 0.04, 0.05},
xticklabels={
  \(\displaystyle {0.01}\),
  \(\displaystyle {0.02}\),
  \(\displaystyle {0.03}\),
  \(\displaystyle {0.04}\),
  \(\displaystyle {0.05}\), 
},
scaled ticks=false,
ymin=0,
ymax=5000,
ylabel style={font=\color{white!15!black}},
ylabel={k-QAoI},
axis background/.style={fill=white},
legend style={at={(0.38,0.30)}, anchor=south west,legend cell align=left, align=left, draw=white!15!black}
]
\addplot [color=orange, line width=1.5pt, mark size=1.5pt, mark=o, mark options={solid, orange}]
  table[row sep=crcr]{%
0.005	6.78659149822351\\
0.01	158.333356642161\\
0.015	188.419995813023\\
0.02	244.643363771423\\
0.025	333.067076155617\\
0.03	456.645974827451\\
0.035	614.85232486133\\
0.04	790.735390611346\\
0.045	982.911988307852\\
0.05	1183.55049410681\\
};
\addlegendentry{\fontsize{8},$\text{CoWu (Theory)}$}

\addplot [color=black, dashed, line width=2.0pt]
  table[row sep=crcr]{%
0.005	29.2258395750804\\
0.01	1265.50514073756\\
0.015	2518.63033363083\\
0.02	3144.85418871071\\
0.025	3521.80223488347\\
0.03	3771.26241163843\\
0.035	3950.3149418398\\
0.04	4084.65008761638\\
0.045	4190.70642205578\\
0.05	4274.71314430574\\
};
\addlegendentry{\fontsize{8},$\text{\gls{rr} (Theory)}$}

\addplot [color=black, line width=1.5pt, only marks, mark size=1.5pt, mark=x, mark options={solid, black}]
  table[row sep=crcr]{%
0.005	28.8702569198748\\
0.01	1278.40303143151\\
0.015	2515.28714193897\\
0.02	3140.64673610659\\
0.025	3525.89097025362\\
0.03	3771.70898751707\\
0.035	3949.60347699255\\
0.04	4101.71879667906\\
0.045	4203.54781321184\\
0.05	4282.33613318336\\
};
\addlegendentry{\fontsize{8},$\text{\gls{rr} (Simulation)}$}

\end{axis}
\end{tikzpicture}%
\caption{Optimal \gls{k-qaoi} of \gls{cowu} against $\alpha$.}
\label{Fig:Optimal_K_QAoI_Against_alpha}
\end{figure}

Figs.~\ref{Fig:zeta_opt_Against_alpha} and~\ref{Fig:Optimal_K_QAoI_Against_alpha} show the optimal timing of \gls{cowu} signal and optimal \gls{k-qaoi} against the value of $\alpha$ for the exponential age in Eq.~\eqref{eq:cost_function}, where we set $N = 100$, $k = 5$, $e_{c} = 0$, $p = 0.0606$, $\Gamma = 1000$, and $V_{\mathrm{th}} = 46$. 
Here, the optimal timing of wake-up signaling is obtained based on the grid search.  
Note that the value of $\alpha$ represents the different speeds of changes of the physical process: smaller (larger) $\alpha$ represents a slowly (fast) changing physical process. From Fig.~\ref{Fig:zeta_opt_Against_alpha}, we can see that, in the range of large $\alpha$, the value of $\zeta_{\mathrm{opt}}$ becomes smaller, in which each node should transmit data later time in order to prevent data staleness at the deadline. This is because earlier transmission of wake-up signal leads to data staleness when the speed of change of the observed physical process is higher.  
On the other hand, when $\alpha$ is smaller, the earlier transmission of \gls{cowu} signal (a larger value of $\zeta$) is the best strategy in order to reduce \gls{k-qaoi}, by which the probability of succeeding in data transmission by the deadline for each wake-up node increases. 
Here, with a smaller range of $\alpha$, e.g. ($\alpha < 0.01$), we can see that the $\zeta_{\mathrm{opt}}$ becomes smaller as the value of $\alpha$ decreases. In this range, thanks to the slowly changing physical process, failure of data transmission does not significantly deteriorate \gls{k-qaoi} as $f_{s}(\Gamma)$ is small. Thus, the sink can transmit a wake-up signal at a later time to collect fresher information from the sensor nodes to reduce  \gls{k-qaoi}. The result of Fig.~\ref{Fig:zeta_opt_Against_alpha} brings us insight into the importance of the timing of wake-up signaling in accordance with the characteristics of physical process observed by each sensor. 

Next, from Fig.~\ref{Fig:Optimal_K_QAoI_Against_alpha}, we can see that \gls{k-qaoi} of both schemes becomes larger as $\alpha$ becomes larger. This is because collected data becomes obsolete faster when the speed of changes of the observed physical process is higher. Focusing on the difference of schemes, from Fig.~\ref{Fig:Optimal_K_QAoI_Against_alpha}, we can see that the performance of \gls{cowu} outperforms \gls{rr} in terms of \gls{k-qaoi}. In \gls{rr}, some nodes transmit data earlier to realize reliable data collection, by which the age of these earlier transmission nodes becomes larger, especially when the value of $\alpha$ is higher. On the other hand, \gls{cowu} can achieve smaller \gls{k-qaoi} because only nodes that observe value equal to or more than $V_{\mathrm{th}}$ wake up and transmit data in the relatively late time, which prevents collected data at the deadline from being obsolete. 

\subsubsection{Impact on the Timing of \gls{cowu} Signaling against $\Gamma$}\label{sec:Investigation_of_Gamma}
\begin{figure}[t]
\centering
%
%
\begin{tikzpicture}

\begin{axis}[%
width=7cm,
height=4cm,
scale only axis,
xmin=1000,
xmax=10000,
xlabel style={font=\color{white!15!black}},
xlabel={$\Gamma$},
xtick={2000, 4000,6000,8000, 10000},
scaled ticks=false,
ymin=150,
ymax=230,
ylabel style={font=\color{white!15!black}},
ylabel={$\zeta{}_{\text{opt}}$},
axis background/.style={fill=white},
legend style={at={(0.3,0.211)}, anchor=south west, legend cell align=left, align=left, draw=white!15!black}
]
\addplot [color=orange, line width=1.5pt, mark size=1.5pt, mark=o, mark options={solid, orange}]
  table[row sep=crcr]{%
1000	170\\
2000	190\\
3000	210\\
4000	210\\
5000	220\\
6000	220\\
7000	220\\
8000	220\\
9000	220\\
10000	220\\
};
\addlegendentry{CoWu (Theory)}

\end{axis}
\end{tikzpicture}%
\caption{Optimal timing of \gls{cowu} signal $\zeta_{\mathrm{opt}}$ against $\Gamma$ (Linear age).}
\label{Fig:zeta_opt_Against_Gamma}
\end{figure}
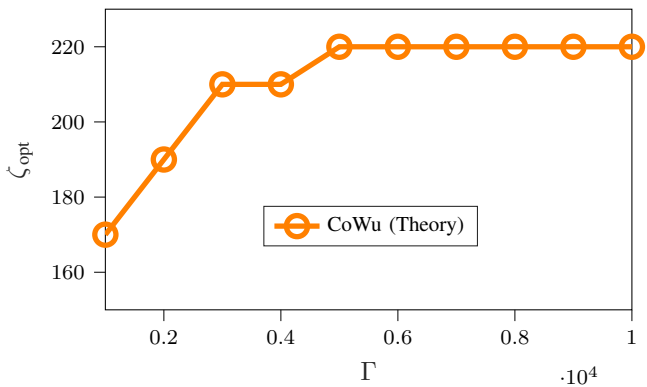

Fig.~\ref{Fig:zeta_opt_Against_Gamma} shows the optimal timing of \gls{cowu} signaling denoted as $\zeta_{\mathrm{opt}}$ against the age penalty $\Gamma$, where we set $N = 100$, $k = 5$, $e_{c} = 0$, $p = 0.0606$, and $V_{\mathrm{th}} = 46$ with linear age scenario of Eq.~\eqref{eq:cost_function}. From Fig.~\ref{Fig:zeta_opt_Against_Gamma}, we can see that an optimal timing of \gls{cowu} signaling $\zeta_{\mathrm{opt}}$ becomes larger as the value of $\Gamma$ increases. Recall that, the value of $\Gamma$ corresponds to the penalty for the data collection failure. If the value of $\Gamma$ is small, the value of the penalty for the data collection failure becomes smaller, which realizes a lower \gls{k-qaoi} of nodes when failing to report by the deadline. In this case, it is desirable for the sink to transmit a \gls{cowu} signal at the time just before the deadline to improve \gls{aoi} at the deadline. On the other hand, the larger $\Gamma$ implies the sink has not received data for a very long period of time, so it can not exploit any temporal knowledge. In this case, reliability for data collection becomes more important, as it gives the sink fresh information on top-$k$ set. That is why the optimal timing of \gls{cowu} signal $\zeta$ becomes larger when the value of $\Gamma$ increases. Finally, it can be observed that the value of $\zeta_{\mathrm{opt}}$ becomes constant after $\Gamma$ exceeds 5,000 because of the constraint of maximum age $A_{\mathrm{max}}$.

\subsection{Effectiveness of \gls{cowu}}

\subsubsection{Investigation of the Impact of Packet Error}\label{sec:Paket_Error}
Fig.~\ref{Fig:k_QAoI_Agaisnt_e_c_1000_5000} shows \gls{k-qaoi} of \gls{cowu} and \gls{rr} against the value of $e_{c}$, where we set $N = 100$, $k = 5$, $V_{\mathrm{th}} = 46$, $\zeta = 150$, and the optimal transmission probability of $p_{\mathrm{opt}}^{*}$  obtained by Eq.~\eqref{eq:dynamic_p}, with linear age scenario in Eq.~\eqref{eq:cost_function}, and evaluate the cases of $\Gamma = 1000$ and $\Gamma = 5000$. 

First, from Fig.~\ref{Fig:k_QAoI_Agaisnt_e_c_1000_5000}, we can see that the results for \gls{cowu} obtained with theoretical analysis coincide with the simulation results, which confirm the validity of our analysis applying optimal transmission probability. Next, from Fig.~\ref{Fig:k_QAoI_Agaisnt_e_c_1000_5000}, we can see that \gls{k-qaoi} of each scheme becomes larger as the value of $e_{c}$ increases. In \gls{cowu}, the node that fails to transmit data due to the single packet error continues to attempt data transmission following $p$-persistent \gls{csma}, which increases the contention time and decreases the probability of succeeding in data transmission by the deadline, which is why \gls{k-qaoi} increases as the value of $e_{c}$ increases. On the other hand, in \gls{rr}, if a single packet error occurs for a time slot, the sink can not retrieve the information of the observation values of this failed packet, by which \gls{k-qaoi} of \gls{rr} becomes larger. Next, focusing on the case where the penalty for the data transmission failure is relatively small, i.e., the case for $\Gamma = 1000$ in Fig.~\ref{Fig:k_QAoI_Agaisnt_e_c_1000_5000}, we can see that \gls{cowu} achieves smaller \gls{k-qaoi} than that of \gls{rr}. This is because, in \gls{cowu}, each node can retransmit data following the operation of $p$-persistent \gls{csma} when experiencing a single packet error, which increases the chance of success in data transmission. On the other hand, in \gls{rr}, as mentioned above, the \gls{k-qaoi} becomes higher by the penalty of $\Gamma$ when the single packet error occurs. Next, focusing on the case for $\Gamma = 5000$ shown in Fig.~\ref{Fig:k_QAoI_Agaisnt_e_c_1000_5000}, first, we can see that the slope of \gls{k-qaoi} of both \gls{cowu} and \gls{rr} is larger than that of the case for $\Gamma = 1000$ shown in Fig.~\ref{Fig:k_QAoI_Agaisnt_e_c_1000_5000}. This is simply because of the increasing penalty value of $\Gamma$ for the data collection failure.  
When the value of $\Gamma$ is higher, it is important for the sink to collect the fresh data from the sensor nodes. In \gls{cowu}, data transmission failure may occur due to the congestion or inappropriate threshold of \gls{cowu}, by which the \gls{k-qaoi} becomes larger, especially when the value of $\Gamma$ is larger. However, even if the value of $\Gamma$ is higher, such as $\Gamma = 5000$, we can see that the  \gls{cowu} can achieve much smaller \gls{k-qaoi} than that of \gls{rr} especially when the value of $e_{c}$ is larger.

 \begin{figure}[t]
\centering
%
%
\definecolor{mycolor1}{rgb}{0.00000,0.44706,0.74118}%
\definecolor{mycolor2}{rgb}{0.00000,0.49804,0.00000}%
\begin{tikzpicture}

\begin{axis}[%
width=7cm,
height=4cm,
scale only axis,
xmin=0,
xmax=0.2,
xlabel style={font=\color{white!15!black}},
xlabel={$\text{e}_{\text{c}}$},
xtick={0, 0.05, 0.1, 0.15, 0.2},
xticklabels={
  \(\displaystyle {0}\),
  \(\displaystyle {0.05}\),
  \(\displaystyle {0.1}\),
  \(\displaystyle {0.15}\),
  \(\displaystyle {0.2}\), 
},
ymin=0,
ymax=4000,
ylabel style={font=\color{white!15!black}},
ylabel={k-QAoI},
axis background/.style={fill=none},
legend style={at={(0.,0.95)}, anchor=north west, legend cell align=left, align=left, legend columns=1, draw=white!15!black}
]
]



\addplot [color=red, line width=1.5pt]
  table[row sep=crcr]{%
0	203.641242398716\\
0.05	214.805252793353\\
0.1	228.652542411656\\
0.15	245.470856444564\\
0.2	265.516280421873\\
};
\addlegendentry{\fontsize{8}, CoWu (Theory,  $\Gamma = 1000$)}

\addplot [color=red, line width=1.5pt, only marks, mark size=1.5pt, mark=star, mark options={solid, red}]
  table[row sep=crcr]{%
0	205.4115\\
0.05	215.518\\
0.1	232.79\\
0.15	247.631\\
0.2	268.031\\
};
\addlegendentry{\fontsize{8}, CoWu (Simulation, $\Gamma = 1000$)}

\addplot [color=mycolor1, dashed, line width=1.5pt, mark size=1.5pt, mark=+, mark options={solid, mycolor1}]
  table[row sep=crcr]{%
0	505\\
0.05	529.7500000000005\\
0.1	554.500000000000\\
0.15	579.250000000000\\
0.2	604\\
};
\addlegendentry{\fontsize{8}, RR-scheduling (Theory,  $\Gamma = 1000$))}



\addplot [color=mycolor2, line width=1.5pt]
  table[row sep=crcr]{%
0	456.070618392672\\
0.05	519.771148291481\\
0.1	598.782153760625\\
0.15	694.745475007217\\
0.2	809.122305936568\\
};
\addlegendentry{\fontsize{8}, CoWu (Theory, $\Gamma = 5000$)}

\addplot [color=mycolor2, line width=1.5pt, only marks, mark size=1.5pt, mark=square, mark options={solid, mycolor2}]
  table[row sep=crcr]{%
0	462.049\\
0.05	526.457\\
0.1	602.893\\
0.15	698.729\\
0.2	842.095\\
};
\addlegendentry{\fontsize{8}, CoWu (Simulation, $\Gamma = 5000$)}

\addplot [color=black, dashed, line width=1.5pt, mark size=1.5pt, mark=triangle, mark options={solid, rotate=180, black}]
 table[row sep=crcr]{%
0	505\\
0.05	729.750000000000\\
0.1	954.500000000000\\
0.15	1179.25000000000\\
0.2	1404\\
};
\addlegendentry{\fontsize{8}, RR-scheduling (Theory,  $\Gamma = 5000$)}
\end{axis}

\end{tikzpicture}%
\caption{\gls{k-qaoi} of \gls{cowu} and \gls{rr} against $e_{c}$ (Linear age), where we apply $\Gamma = 1,000$ and $\Gamma = 5,000$.}
\label{Fig:k_QAoI_Agaisnt_e_c_1000_5000}
\end{figure}

\subsubsection{Minimum Energy Consumption}\label{Sec:Minimum_Ene}

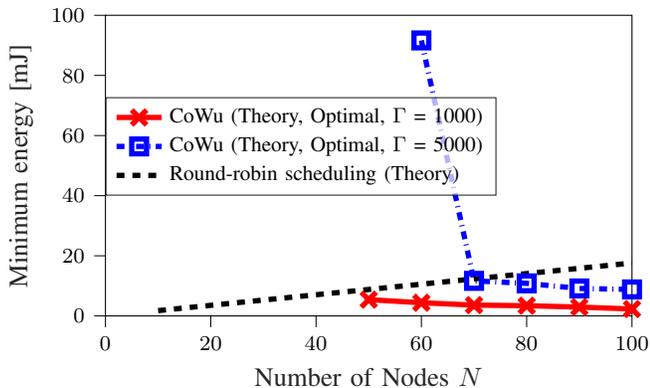
\begin{figure}[t]
\centering
%
%
\begin{tikzpicture}

\begin{axis}[%
width=7cm,
height=4cm,
scale only axis,
unbounded coords=jump,
xmin=0,
xmax=100,
xlabel style={font=\color{white!15!black}},
xlabel={Number of Nodes $N$},
ymin=0,
ymax=100,
ylabel style={font=\color{white!15!black}},
ylabel={Minimum energy [mJ]},
axis background/.style={fill=white},
legend style={at={(0,0.4)}, anchor=south west, legend cell align=left, align=left, draw=white!15!black}
]



\addplot [color=red, line width=1.5pt, mark size=1.5pt, mark=o, mark options={solid, red}]
  table[row sep=crcr]{%
20	nan\\
40	7.67703394629583\\
60	4.38451666335767\\
80	3.39652505014906\\
100	2.29732381462177\\
};
\addlegendentry{CoWu (Theory)~$\Gamma = 1,000$}

\addplot [color=magenta, dotted, line width=1.5pt, mark size=1.5pt, mark=o, mark options={solid, magenta}]
  table[row sep=crcr]{%
20	nan\\
40	nan\\
60	11.5155904199404\\
80	9.01283568493781\\
100	8.81563222347095\\
};
\addlegendentry{CoWu (Theory)~$\Gamma = 5,000$}

\addplot [color=black, line width=1.5pt]
  table[row sep=crcr]{%
10	1.76\\
20	3.52\\
30	5.28\\
40	7.04\\
50	8.8\\
60	10.56\\
70	12.32\\
80	14.08\\
90	15.84\\
100	17.6\\
};
\addlegendentry{RR-scheduling (Theory)}

\end{axis}

\end{tikzpicture}%
\caption{Total energy consumption against the number of nodes for \gls{cowu} and \gls{rr} (Linear age).}
\label{Fig:Minimum_ene}
\end{figure}

In order to compare the performance of \gls{cowu} and \gls{rr}, in terms of minimum energy consumption under the \gls{k-qaoi} constraint, we obtain the optimal parameter set of $\{V_{\mathrm{th}}, \zeta\}$ in Eq.~\eqref{eq:min_set_ene}, where $\gamma_{\mathrm{th}}$ is set to the \gls{k-qaoi} of \gls{rr}, $\bar{\Delta}_{k}^{\mathrm{RR}}(N)$. Fig.~\ref{Fig:Minimum_ene} shows the minimum total energy consumption of \gls{cowu} obtained by Eq.~\eqref{eq:min_set_ene} with \gls{rr} against $N$, where we set $e_{c} = 0$, $k = 5$ and $\Gamma$ = 1000, 5000 for the linear age case in Eq.~\eqref{eq:cost_function}. In Fig.~\ref{Fig:Minimum_ene}, total energy consumption of \gls{cowu} for the number of nodes or less than 40 is not plotted. This is because \gls{cowu} can not realize smaller \gls{k-qaoi} than \gls{rr} over this range of number of nodes. However, we can see that as the number of nodes becomes larger, \gls{cowu} can satisfy the constraint of \gls{k-qaoi}, in which minimum total energy consumption of \gls{cowu} becomes smaller as $N$ increases. This is because as the number of nodes increases, the \gls{k-qaoi} of \gls{rr} becomes larger, by which the sink can select the appropriate parameter $\zeta$ with a relatively larger threshold of $V_{\mathrm{th}}$ for \gls{cowu} signaling under the constraint of \gls{k-qaoi} of \gls{rr}. Further, from Fig.~\ref{Fig:Minimum_ene}, we can also see that the minimum total energy consumption of \gls{cowu} becomes larger as $\Gamma$ increases. This is because, as $\Gamma$ increases, the sink is required to set a relatively smaller threshold of $V_{\mathrm{th}}$ in order to avoid the increase of \gls{k-qaoi} by the penalty of $\Gamma$, by which total energy consumption of sensor nodes increases. 
Finally, we can see that \gls{cowu} can achieve smaller energy consumption for larger number of nodes while satisfying the constraint of \gls{k-qaoi}, thanks to the suppression of wake-up of nodes that observe unrelated data for the top-$k$ query. From this result, we can clearly see the effectiveness of \gls{cowu} against \gls{rr} in terms of \gls{k-qaoi} and total energy consumption.

\subsubsection{Maximum $k$}

This section investigates the maximum $k$ as formulated in Eq.~\eqref{eq:maximum_k_3} for the four types of settings: 
\begin{itemize}
\item{Setting 1: We assume linear age, $\Gamma = 1000$, and collision channel, i.e., $e_{c} = 0$.}
\item{Setting 2: We assume linear age, $\Gamma = 1000$, and collision channel and erasure channel with the constant error probability $e_{c} = 0.1$.}
\item{Setting 3: We assume linear age, relatively large age-penalty $\Gamma = 5000$, and collision channel i.e., $e_{c} = 0$.}
\item{Setting 4: We assume exponential age with $\alpha = 0.02$, $\Gamma = 1000$, and collision channel i.e., $e_{c} = 0$.}
\end{itemize}
In this analysis, $\gamma_{\mathrm{th}}$ and $\epsilon_{\mathrm{th}}$ in Eq.~\eqref{eq:maximum_k_3} are set to $\bar{\Delta}_{k}^{\mathrm{RR}}(N)$ in Eq.~\eqref{eq:RR_k_QAoI} and  $E_{\mathrm{total}}^{\mathrm{RR}} (N)$ in Eq.~\eqref{eq:RR_ene}, respectively. With this setting, we can investigate the maximum value of $k$, under which the \gls{cowu} outperforms \gls{rr} in terms of total energy consumption and \gls{k-qaoi}.

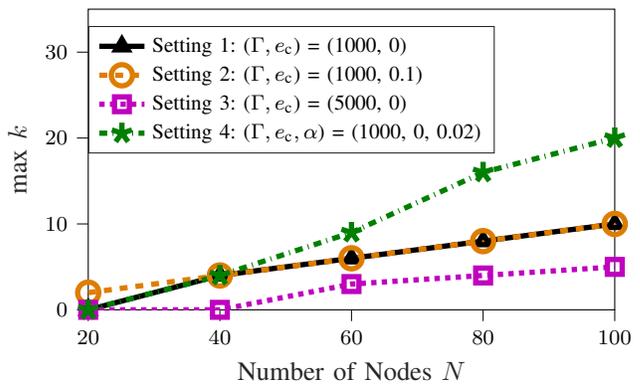
\begin{figure}[t]
\centering
%
%
\definecolor{mycolor1}{rgb}{0.87059,0.49020,0.00000}%
\definecolor{mycolor2}{rgb}{0.74902,0.00000,0.74902}%
\definecolor{mycolor3}{rgb}{0.00000,0.49804,0.00000}%
\begin{tikzpicture}

\begin{axis}[%
width=7cm,
height=4cm,
scale only axis,
xmin=20,
xmax=100,
xlabel style={font=\color{white!15!black}},
xlabel={Number of Nodes $N$},
ymin=0,
ymax=50,
ylabel style={font=\color{white!15!black}},
ylabel={max $k$},
axis background/.style={fill=white},
legend style={at={(0,0.52)}, anchor=south west, legend cell align=left, align=left, draw=white!15!black}
]

\addplot [color=black, line width=1.5pt, mark size=1.5pt, mark=triangle, mark options={solid, black}]
  table[row sep=crcr]{%
20	0\\
40	4\\
60	6\\
80	12\\
100	10\\
};
\addlegendentry{Setting 1: ($\Gamma, e_{\text{c}}$) $=$ (1000, 0)}

\addplot [color=mycolor1, dashed, line width=1.5pt, mark size=1.5pt, mark=o, mark options={solid, mycolor1}]
  table[row sep=crcr]{%
20	2\\
40	4\\
60	6\\
80	8\\
100	10\\
};
\addlegendentry{Setting 2: ($\Gamma, e_{\text{c}}$) $=$ (1000, 0.1)}

\addplot [color=mycolor2, dotted, line width=1.5pt, mark size=1.5pt, mark=square, mark options={solid, mycolor2}]
  table[row sep=crcr]{%
20	0\\
40	0\\
60	3\\
80	4\\
100	5\\
};
\addlegendentry{Setting 3: ($\Gamma, e_{\text{c}}$) $=$ (5000, 0)}

\addplot [color=mycolor3, dashdotted, line width=1.5pt, mark size=1.5pt, mark=star, mark options={solid, mycolor3}]
  table[row sep=crcr]{%
20	0\\
40	4\\
60	9\\
80	20\\
100	20\\
};
\addlegendentry{Setting 4: ($\Gamma, e_{\text{c}}, \alpha$) $=$ (1000, 0, 0.02)}

\end{axis}
\end{tikzpicture}%
\caption{Maximum $k$ for \gls{cowu} against the number of nodes with different parameters settings with $p_{\mathrm{opt}}^{*}$ in Eq.~\eqref{eq:dynamic_p}.}
\label{Fig:maxk_dynamic_p}
\end{figure}

Fig.~\ref{Fig:maxk_dynamic_p} shows the results of maximum $k$ against the different parameter settings, where we adopt the optimal transmission probability $p_{\mathrm{opt}}^{*}$ obtained in Eq.~\eqref{eq:dynamic_p}. First, we can see that the value of $k$ becomes larger as the number of nodes increases.  In \gls{rr}, data staleness for the nodes that transmit data very early time can not be avoided to realize reliable data transmission, which increases the average \gls{k-qaoi} and its negative effect becomes larger as the number of nodes increases. On the other hand, \gls{cowu} achieves both high information freshness and high energy efficiency by carefully tuning the parameters $\{\zeta$, $V_{\mathrm{th}}\}$, especially when the value of $k$ is small. 

Next, focusing on the different settings, we can see that setting 3 achieves the lowest maximum $k$ than other settings. When the penalty for data transmission failure is large, which corresponds to settings 3, the sink needs to set a relatively small threshold $V_{\mathrm{th}}$ of \gls{cowu} and transmit \gls{cowu} signal earlier time in order to alleviate the negative effect for the data collection failure, which increases the energy consumption and \gls{k-qaoi}. Comparing the results of setting-1 and 3, we can see that the maximum $k$ of setting-3 is always smaller than setting-1, depicted by a black line. This is because, as the value of $\Gamma$ becomes larger, \gls{k-qaoi} of \gls{cowu} becomes larger as discussed in Sec.~\ref{sec:Investigation_of_Gamma}, while total energy consumption and \gls{k-qaoi} of \gls{rr} only depends on the number of nodes $N$, when $e_{c} = 0$. 

Furthermore, we can see that the maximum $k$ of setting 2 is equal to or lower than that of setting 1 when $N >20$. In \gls{cowu}, a longer contention period is required when a single packet error happens, which increases the \gls{k-qaoi} and total energy consumption. That is why the maximum $k$ of setting 2 is equal to or lower than that of setting 1. On the other hand, when the number of nodes is 20, maximum $k$ of setting 2 becomes larger than that of setting 1. This is because when the single packet error occurs, \gls{k-qaoi} of \gls{rr} becomes larger by the penalty of $\Gamma$, as it decreases the reliability of data collection.

Finally, from Fig.~\ref{Fig:maxk_dynamic_p}, we can see that the maximum $k$ of settings 4, in which we apply the exponential age (loss) function, is much larger than other settings. This is mainly because the \gls{k-qaoi} of \gls{rr} becomes significantly larger due to the requirement to transmit data at an earlier time, whose age increases faster than the case of linear age. On the other hand, in \gls{cowu}, we can suppress the increasing value of loss for the case of exponential age in Eq.~\eqref{eq:cost_function} by activating only the subset of nodes whose observations are highly likely to belong to the top-$k$ set at $T$ by sending wake-up signal just before the deadline.

From these observations, we can conclude that \gls{cowu} outperforms \gls{rr} in the context of timely top-$k$ data collection in the \gls{iiot} scenario in terms of total energy consumption and \gls{k-qaoi} when the number of nodes equal to or larger than 60. We also calculate the ratio of the maximal value of $k$ against $N$, which we call the maximum $k$-$N$ ratio as in~\cite{TGCN_Content}, based on the results obtained in Fig.~\ref{Fig:maxk_dynamic_p}, which shows the maximum region that the \gls{cowu} outperforms the \gls{rr} in terms of \gls{k-qaoi} and total energy consumption. With our calculation, we confirmed that the maximum $k$-$N$ ratio reaches 0.25.

Finally, this article reveals the effectiveness of \gls{cowu} under the assumption that the sink has perfect knowledge of the parameter of the physical process, such as $\alpha$, by which the sink can select the optimal parameters for \gls{cowu} signal, i.e., $\zeta$ and $V_{\mathrm{th}}$. It is worth emphasizing that, in practice, the parameter of $\{\zeta, V_{\mathrm{th}}\}$ should be carefully chosen considering the uncertainty of knowledge/statistics (e.g., $\alpha$), as it directly affects the performance of \gls{cowu}.

\section{Conclusions}\label{Sec:conc}
In this article, we have applied \gls{cowu} in a scenario where the timeliness of top-$k$ data at the deadline is crucial to realize high information freshness of top-$k$ data and high energy efficiency of sensor nodes. In order to investigate the performance of \gls{cowu}, we have introduced a new metric called \gls{k-qaoi} and derived equations of \gls{k-qaoi} and total energy consumption assuming $p$-persistent \gls{csma} as a \gls{mac} protocol. The performance of \gls{cowu} in the timely top-$k$ data collection was thoroughly investigated through a wide range of parameters, which clarified the importance of timing of \gls{cowu} signaling and the threshold of \gls{cowu}, considering the required size of top-$k$ data. With numerical evaluations, we have revealed the region and scenario where \gls{cowu} outperforms \gls{rr} in terms of \gls{k-qaoi} and total energy consumption.

Possible avenues for future work include designing a wake-up control introducing the inference mechanism of the physical process to decide the optimal timing of sampling, wake-up, and data transmission for sensor nodes. Designing the wake-up control aiming for the improvement of the model accuracy on the physical process is also an interesting direction. 
\ifCLASSOPTIONcaptionsoff
  \newpage
\fi

\bibliographystyle{IEEEtran}
\bibliography{IEEEabrv,Ref}

\end{document}